\documentclass[prb,twocolumn,superscriptaddress,showpacs,amsmath,amssymb]{revtex4}

\usepackage{graphicx}
\usepackage{latexsym}
\usepackage{amsmath}
\usepackage{amssymb}
\usepackage{amsfonts}
\usepackage{color}
\usepackage{bm}
\usepackage{bbm}

\usepackage{verbatim}

\begin{document}

\title{Large enhancement of spin pumping due to the surface bound states in 
normal metal/superconductor structures}

 \author{M.A.~Silaev}
 \affiliation{Department of
Physics and Nanoscience Center, University of Jyv\"askyl\"a, P.O.
Box 35 (YFL), FI-40014 University of Jyv\"askyl\"a, Finland}
\affiliation{Moscow Institute of Physics and Technology, Dolgoprudny, 141700 Russia}
\affiliation{Institute for Physics of Microstructures, Russian Academy of Sciences, 603950 Nizhny Novgorod, GSP-105, Russia}

 \begin{abstract}
 We show that the spin pumping from ferromagnetic insulator into the adjacent metallic spin sink can be strongly stimulated by the superconducting correlations.  
 The key physical mechanism responsible for this effect  is the presence of  quasiparticle surface states  at the ferromagnetic insulator/superconductor  interface. 
We consider the minimal model when these states  appear because of the suppressed pairing constant within the  interfacial normal layer. 
For thin normal layers  we obtain   a strongly  peaked 
temperature dependence  of the Gilbert damping coefficient which has been recently observed in such systems. For thicker normal layers the Gilbert damping monotonically increases  down to the temperatures much smaller than the critical one. The suggested model paves the way to controlling the temperature dependence of the spin pumping  by fabricating 
hybrid normal metal/superconductor spin sinks.   
 \end{abstract}

 \maketitle

{\bf Introduction}
 \begin{figure}
 \centerline{
 $  \begin{array}{c}
 \includegraphics[width=1.0\linewidth]
 {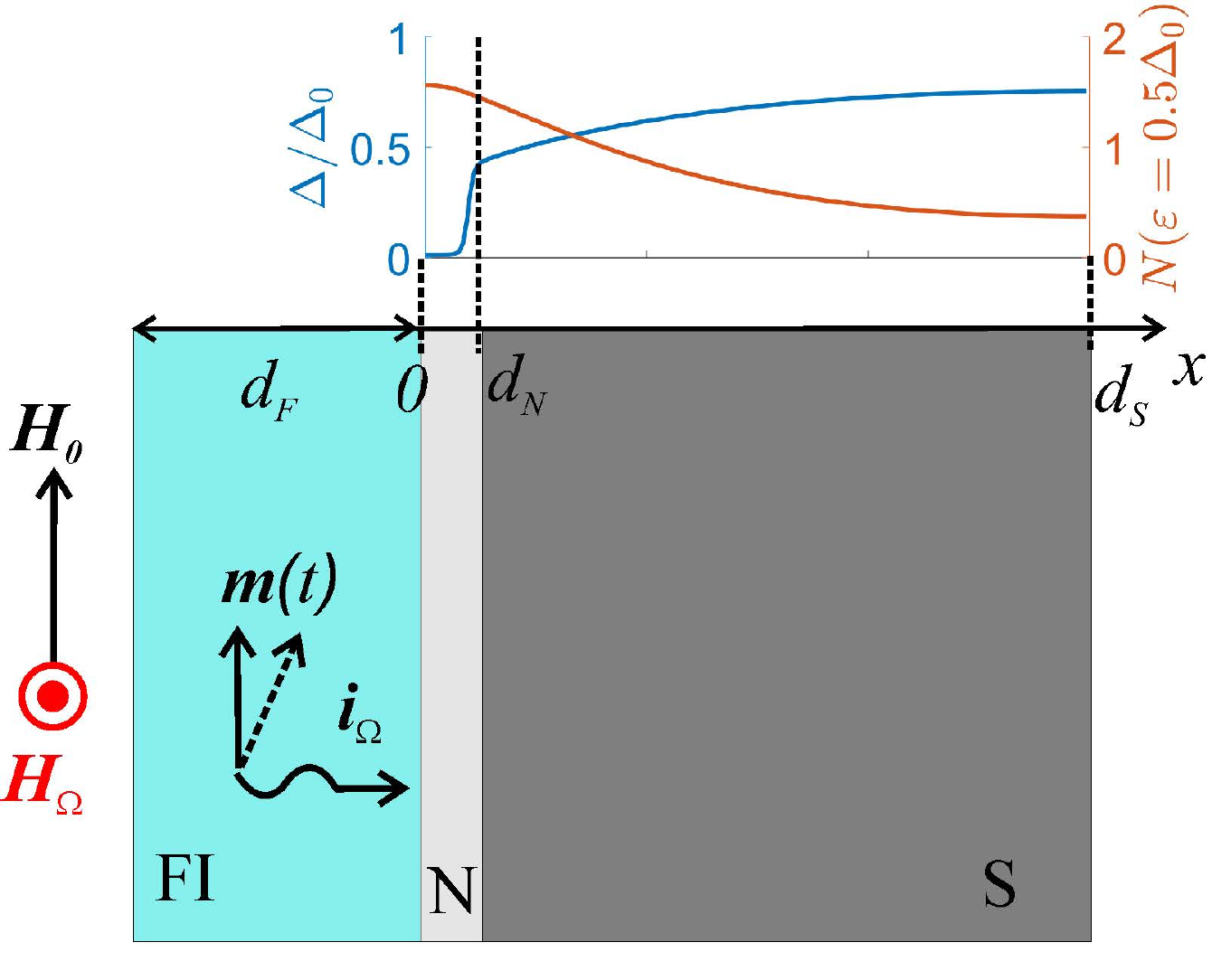} 
  \end{array}$}
  \caption{\label{Fig:1}
  Schematic setup of the ferromagnetic insulator (FI) film  with the adjacent metallic spin sink  consisting of 
  of normal (N) and superconducting (S) layers. 
 The constant external magnetic field is $ H_0 \bm x$.  
 The magnetization precession $\bm m(t)$ is driven by the external magnetic field $ H_\Omega e^{i\Omega t} \bm y$. It generates spin current 
 $\bm i_\Omega$  pumped from F to the spin sink. 
 Upper panel shows the coordinate dependencies of the order parameter $\Delta (x)$ and local density of states $N(x)$ at the
 energy $\varepsilon =0.5\Delta_0$ for $d_N = 0.2\xi_0$, $d_S =3 \xi_0$, $T=0.7 T_{c}$. 
    }
 \end{figure}
Spin transport and spin dynamics in superconductors have attracted significant attention recently\cite{Linder2015,RevModPhys.90.041001, han2019spin,quay2018out,ohnishi2020spin,beckmann2016spin,Eschrig2015a}. 
Quite interesting experimental results have been obtained 
for the spin pumping effects \cite{bell2008spin,jeon2019abrikosov,PhysRevB.99.024507,PhysRevApplied.11.014061,Jeon2018,yao2018probe,li2018possible,
zhao2020exploring,golovchanskiy2020magnetization,jeon2020tunable,2004.09467} which in general play the central role in  spintronics \cite{brataas2002spin,tserkovnyak2002enhanced,RevModPhys.77.1375}. 
It was found that superconducting correlations can lead either to the significant suppression \cite{bell2008spin} or to the significant enhancement \cite{yao2018probe,jeon2019abrikosov,PhysRevB.99.024507,PhysRevApplied.11.014061,Jeon2018, jeon2020tunable} of Gilbert damping (GD)  coefficient  in  systems consisting of  superconducting and ferromagnetic layers, such as in the generic example shown in Fig.\ref{Fig:1}. 
The basic mechanism for changing  GD in such systems is the spin pumping effect \cite{}. This mechanism is based on the 
spin angular momentum transfer from the ferromagnet into the the adjacent metallic film via the pumped spin current $\bm i (t)$ generated by the time-dependent magnetization $\bm m (t)$. 
The spin relaxation  in the  metallic spin sink leads to the damping-like spin torque and modifies the effective GD coefficient  of the system. 

In this way the 
suppression of GD with decreasing temperature $T<T_c$ in systems with superconducting spin sink\cite{bell2008spin} can be qualitatively understood as resulting from the the freezing out of quasiparticles in the superconductor\cite{morten2008proximity}.   
However, the strong increase of GD with lowering temperature \cite{yao2018probe,jeon2019abrikosov,PhysRevB.99.024507,PhysRevApplied.11.014061,Jeon2018, jeon2020tunable} seems to be counter-intuitive and its understanding requires further theoretical efforts.
 
In ferromagnetic insulator (FI) /superconductor (S) bilayers  GdN/NbN    the peaked behaviour of GD as a function of temperature has been observed \cite{yao2018probe}.  The maximal  GD reached at about $ T\approx 0.7 T_c$ is several times larger than in the  normal state  $\delta\alpha /\delta\alpha_N \sim  2-3 $, where $\delta\alpha$ is the spin-pumping related change of GD. 
Because of the several reasons such behaviour cannot be explained\cite{2007.04372}  by the coherence peak of spin susceptibility in homogeneous superconductors \cite{TinkhamBook}. 
First, 
this peak occurs at $T\approx 0.9 T_c$  and 
 for the realistic values of the Dynes parameter\cite{dynes84} $\Gamma \approx 0.1 T_c$ in NbN its magnitude is\cite{2007.04372}  $\delta\alpha /\delta\alpha_N \sim  0.2-0.3 $. Such behaviour is typical for the  line widths of nuclear magnetic resonance\cite{hebel1959nuclear, masuda1962nuclear} and electronic paramagnetic resonance\cite{tagirov1987spin}   in superconductors. 
 It is clearly different from the observed  behaviour of GD in FI/S systems \cite{yao2018probe} which has  an order of magnitude larger peak  $\delta\alpha /\delta\alpha_N \sim  2-3 $  at significantly lower temperatures  $T\approx 0.7 T_c$.

\begin{figure}
 \centerline{
 $  \begin{array}{c}
 \includegraphics[width=0.55\linewidth]
 {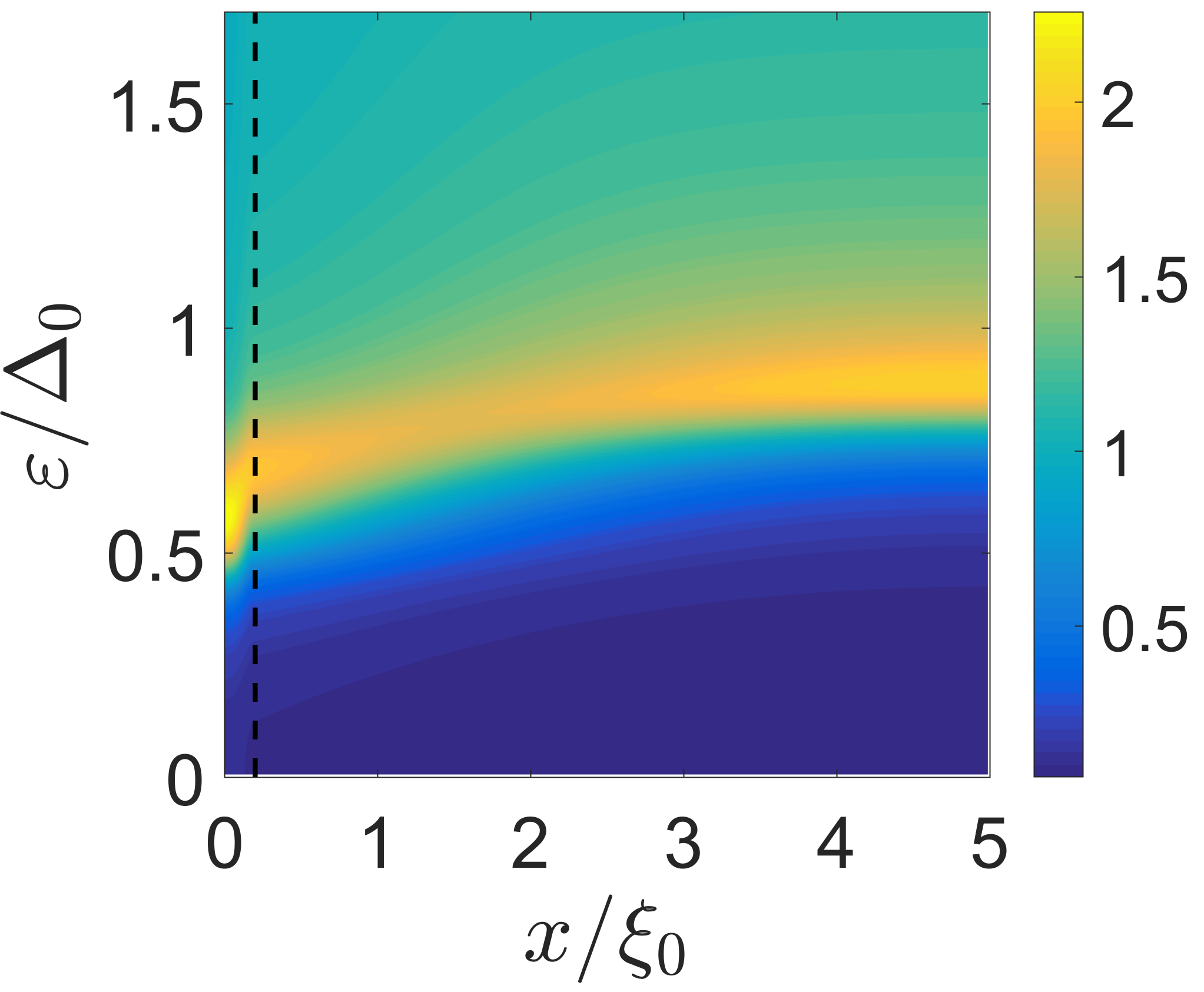}
 \put (-43,95) 
   { \large{\color{black} (a)} }
     \includegraphics[width=0.45\linewidth]
 {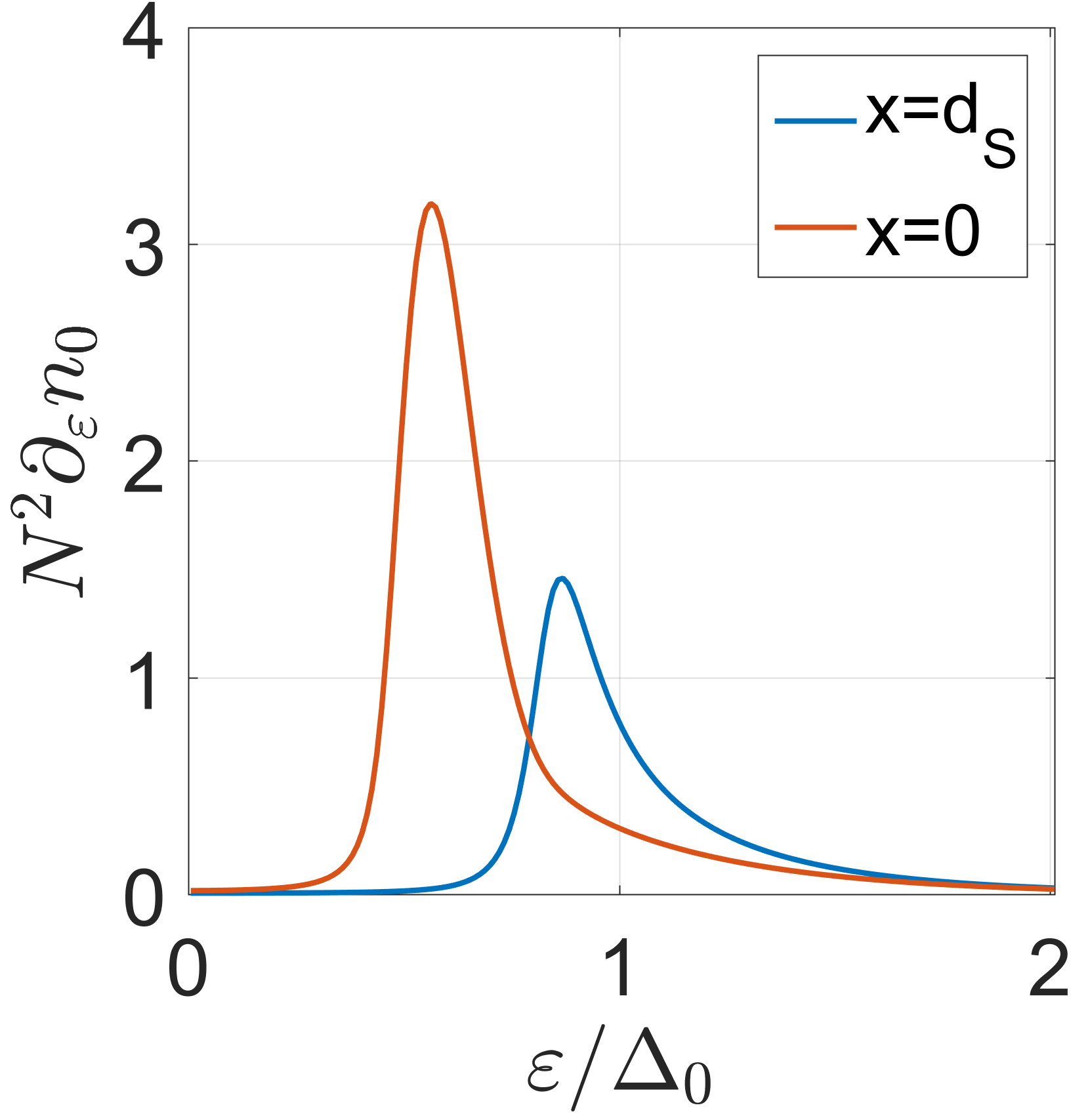} 
  \put (-53,95) 
  { \large{\color{black} (b)} }
   \\
   \includegraphics[width=0.55\linewidth]
 {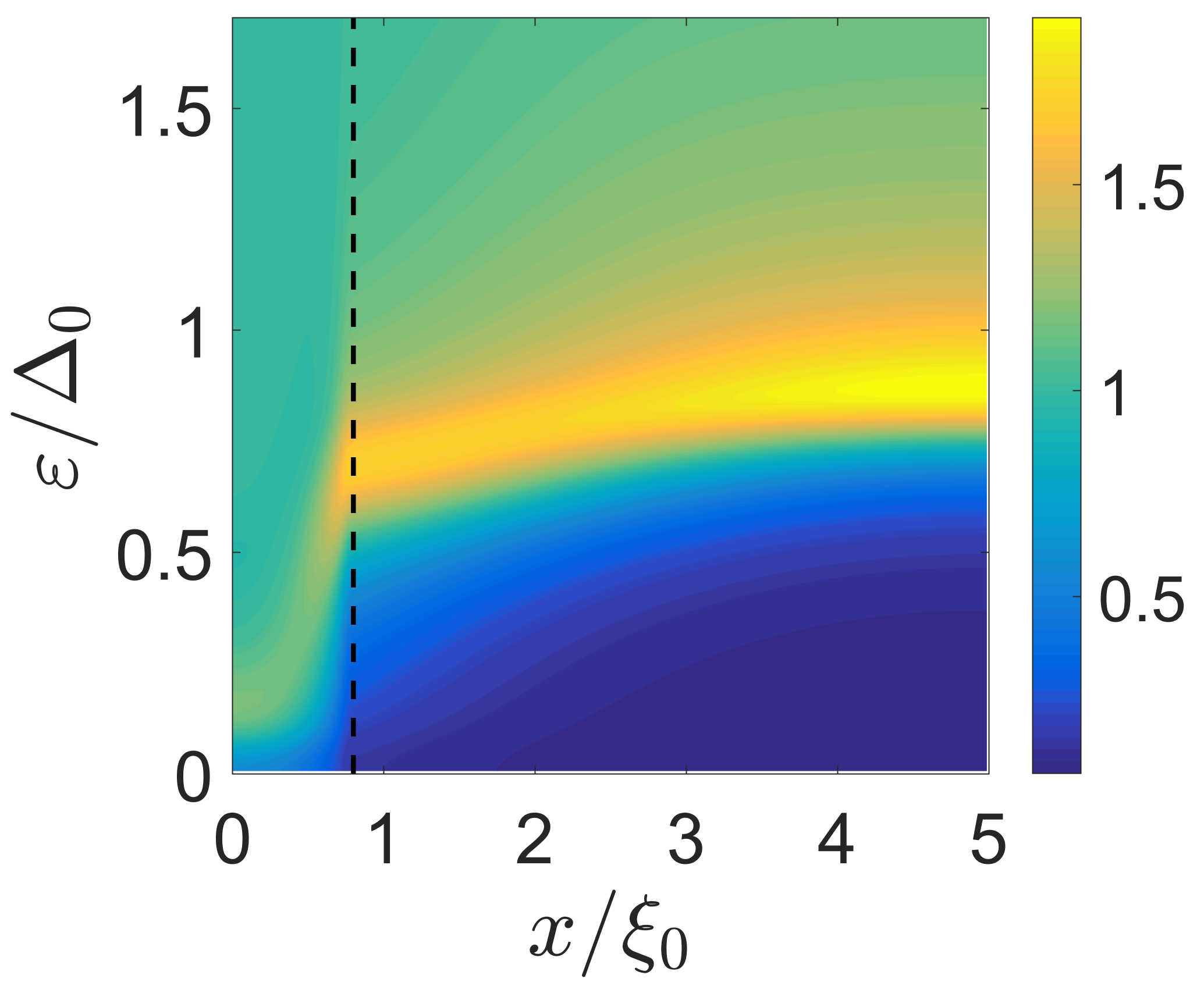}
 \put (-40,90)  
   {\large{\color{black} (c)} }
    \includegraphics[width=0.45\linewidth]
 {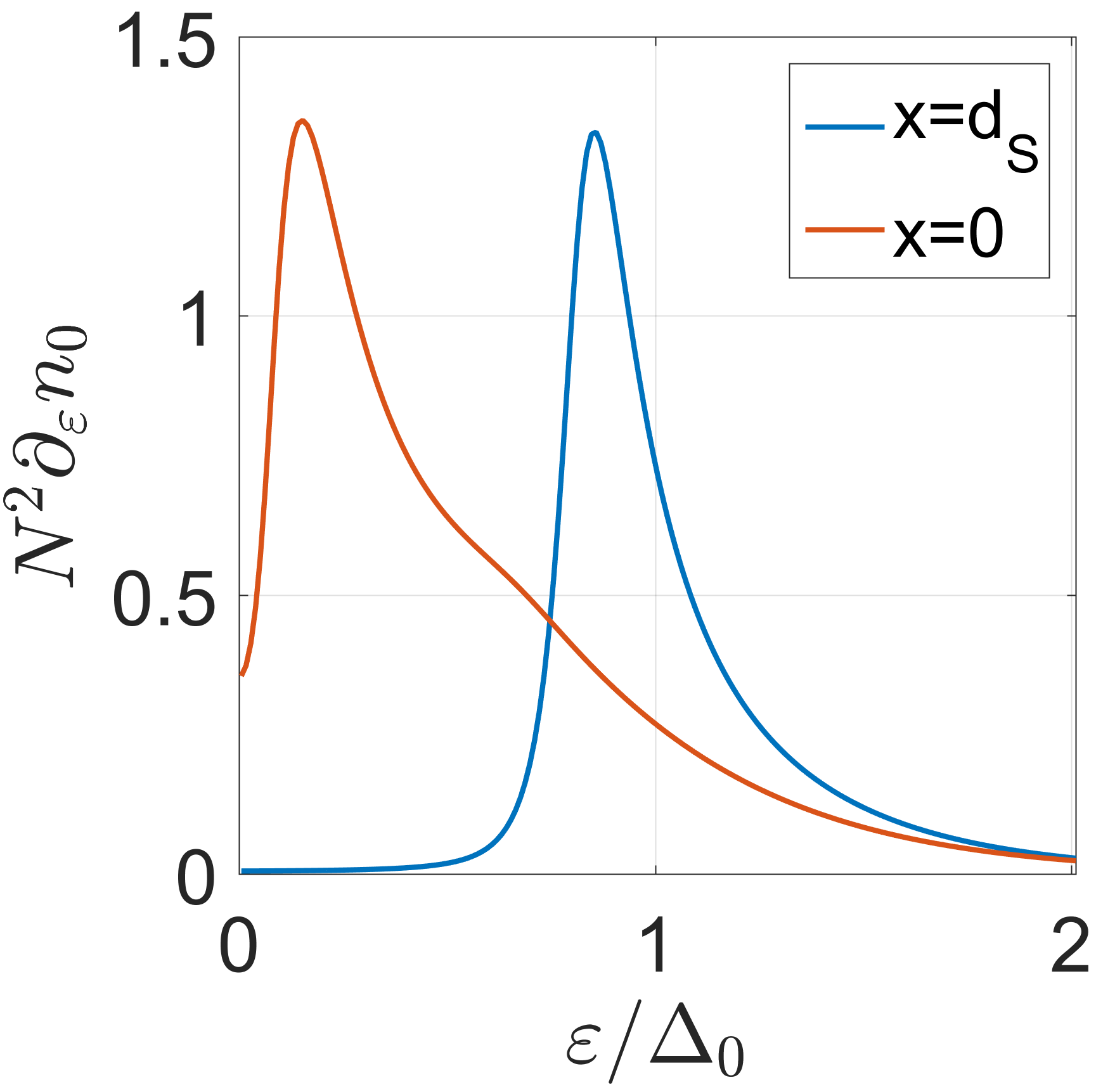} 
  \put  (-47,90) 
  {\large {\color{black} (d)} }
  \end{array}$}
  \caption{\label{Fig:2}
(a,c)  Density of states profile $N(\varepsilon,x)$ in the N/S structure. 
The position of N/S boundary shown by the dashed line
 is at (a) $d_N=0.2\xi_0$ and (c) $d_N=0.8\xi_0$.
 $T=0.7 T_{c0}$, $\Gamma=0.1 T_{c0}$, $d_S=5\xi_0$, $D_N=0.05 D_S$.
 Plots for other $d_S$ are shown in Appendix\cite{SupplMat}.   
  (b,d) Magnon absorption probability $P_m (\varepsilon) = \Omega \partial_\varepsilon n_0 N^2$  for the frequency $\Omega=0.02 T_{c0}$, 
Red and blue curves are taken at  $x=0$ and $x=d_S$, respectively.
Parameters are the same as in (a,b).  
  }
 \end{figure}

In this Letter we suggest a minimal theoretical model which explains the large  enhancement of GD in FI/S structures.
The key physical mechanism  responsible for this effect is the existence of quasiparticle states localized at the FI/S interface. Such states appear due to the suppressed pairing within the interfacial normal layer\cite{golubov1988theoretical,
golubov1994proximity,
golubov1995proximity,
gurevich2017surface} (N) as illustrated in Fig.\ref{Fig:1}.
Shown on top of the Fig.\ref{Fig:1} are the spatial profiles of the order parameter $\Delta(x)$ and the local density of states (DOS) $N(x)$ at the subgap energy
$\varepsilon=0.5\Delta_0$, where 
 $\Delta_0$ is the bulk energy gap at $T=0$. 
 The overall N/S film  thickness is $d_S=3\xi_0$, where 
 $\xi_0 = \sqrt{D_S/T_{c0}}$ is the coherence length,  $D_S$ is the diffusion constant in S,   $T_{c0}$ is the bulk critical temperature.
Near the interface at $x=0$ the DOS is enhanced due to the subgap  quasiparticle  states  which are formed in the N/S structure \cite{zhou1998density, gupta2004anomalous, 
le2008phase,fominov2001superconductive} and occupy the certain energy interval
between  the bulk gap and Thouless energy $D_N/d_N^2$ where $D_N$ is the diffusion coefficient and $d_N$ is the thickness of N. The existence of surface bound states in N/S structures is demonstrated\cite{SupplMat} in Fig.\ref{Fig:2}a,c where the 
  $N(x,\varepsilon)$
 profiles are shown to have a maximum at $x=0$ and energies which depend on $d_N$. 
 The order parameter and DOS in Figs.\ref{Fig:1},\ref{Fig:2}  are calculated within the Usadel theory \cite{usadel1970generalized}
as explained below. In Fig.\ref{Fig:1} we choose identical diffusion coefficient in N and S layers $D_N=D_S=D$ while in Fig.\ref{Fig:2}
$D_N=0.05D_S$.

  At low frequencies 
$\Omega\ll \Delta_0$ 
the DOS enhancement leads to the increased  probability of the
 magnon absorption by conductivity electrons in the N/S layer. 
Qualitatively, at a given energy level this probability is
 determined by number of available states for transition $N(\varepsilon)N(\varepsilon+\Omega) \approx N^2(\varepsilon)$
 and the difference of occupation numbers  
 $n_0(\varepsilon+\Omega ) -n_0(\varepsilon) \approx \Omega \partial_\varepsilon n_0$
  where $n_0(\varepsilon) =\tanh (\varepsilon/2T)$ is the equilibrium distribution function. The product of these factors leads to the 
  energy-resolved  magnon absorption probability $P_m =\Omega N^2 \partial_\varepsilon n_0 $. 
  In Fig.\ref{Fig:2}b,d one can that of $P_m(\varepsilon)$
   at $T=0.7 T_{c0}$ 
   is enhanced  at the  boundary of  N layer $x=0$ (red curves) as compared to $x=d_S$ (blue curves).
      Besides that, the localization of surface states is qualitatively equivalent to the decrease of the spin sink volume which and the corresponding increase of the non-equilibrium spin polarization.    As we show by an exact calculation below these mechanisms lead to the large enhancement of spin pumping in the N/S films. 

 
Interestingly, besides explaining the large peak of the spin pumping for $d_N\ll \xi_0$ the  model described above yields also the qualitatively different regime with almost monotonic increase of GD down to the temperatures $T\ll T_c$. This behaviour is obtained for $d_N\sim \xi_0$ when the bound states are pushed down to lower energies as shown in Fig.\ref{Fig:2}c and the absorption probability us enhanced for quasiparticles with $\varepsilon\ll\Delta_0$ which are not frozen out down to the significantly low temperatures  determined by the Thouless energy $T_{th} \approx D_N/d_N^2 $. 
Similar behaviour of GD  has been observed experimentally in
Py/Nb/Pt superconducting heterostructures  \cite{Jeon2018, jeon2020tunable}, although its physical origin can be different.

{\bf Model of spin pumping }
     To quantify the spin pumping effect we consider the microscopic model of the spin-dependent scattering of electrons 
      at the FI interface \cite{Tokuyasu1988,millis1988quasiclassical,SupplMat}.  
      As we show below, it formally yields the spin current   identical to the 
      one given by the  interfacial exchange interaction between the localized spins in FI and conduction elections in the adjacent metal\cite{ohnuma2014enhanced}. 
     Within this model the local spin polarization close to the interface $\bm S(t)$ acts as effective field for the localized magnetic moments. 
     This process can be taken into account by introducing the additional term $\bm i(t)$ into the 
     Landau-Lifshitz-Gilber equation 
 \begin{align} \label{Eq:LLG}
 & (1 + \alpha \bm m \times )\partial_t\bm m + 
 \gamma\bm m\times \bm H_{eff} =  \bm i/S_{F0}d_F 
  \\
   \label{Eq:SpinCurrentOn}
   & \bm i (t)= J_{sd}
  \bm S (t)\times \bm m (t) 
 \end{align}
 Here $S_{F0}$ is the equilibrium spin density in F, $d_F$ is the F film thickness, $\bm H_{eff}$ is the effective field and $\alpha$
 is the intrinsic Gilbert damping coefficient.   The term $\bm i (t)$ can be interpreted as the spin current between FI and metal. 

    To calculate $\bm S (t)$ we need to find the spin response of the superconductor to the interfacial exchange field. 
 In the linear regime it is given by 
 \begin{align} \label{Eq:SOmegaChiM}
 \bm S_{\Omega} = \nu h_{eff} \chi_m \bm m_\Omega  
 \end{align}
 where we introduce the effective exchange field 
 $h_{eff} = J_{sd}/d_S $, normal metal DOS at the Fermi level $\nu$  and the local spin susceptibility $\chi_m$.

   The spin-pumping related change of the GD is determined by the dissipative part of the susceptibility
 \begin{align} \label{Eq:GDFiniteThickness}
 \delta\alpha = C  T_{c0} {\rm Im}\chi_m/\Omega 
  \end{align} 
 where the dimensionless coefficient determining the coupling strength between the FI and metallic films is \cite{2007.04372}
  \begin{align}  \label{Eq:CouplingParameter}
 & C =  \frac{h_{eff}}{T_{c0}} \frac{\nu h_{eff}}{S_{F0}}  
  \frac{d_S}{d_F}  
 \end{align} 
 From there one can see that since $h_{eff}\propto 1/d_S^2 $ the coupling coefficient is  
 $C \propto 1/d_S$. Localization of surface states provides the effective decrease of $d_S$ which leads to the increase of $C$ and the  spin response.

 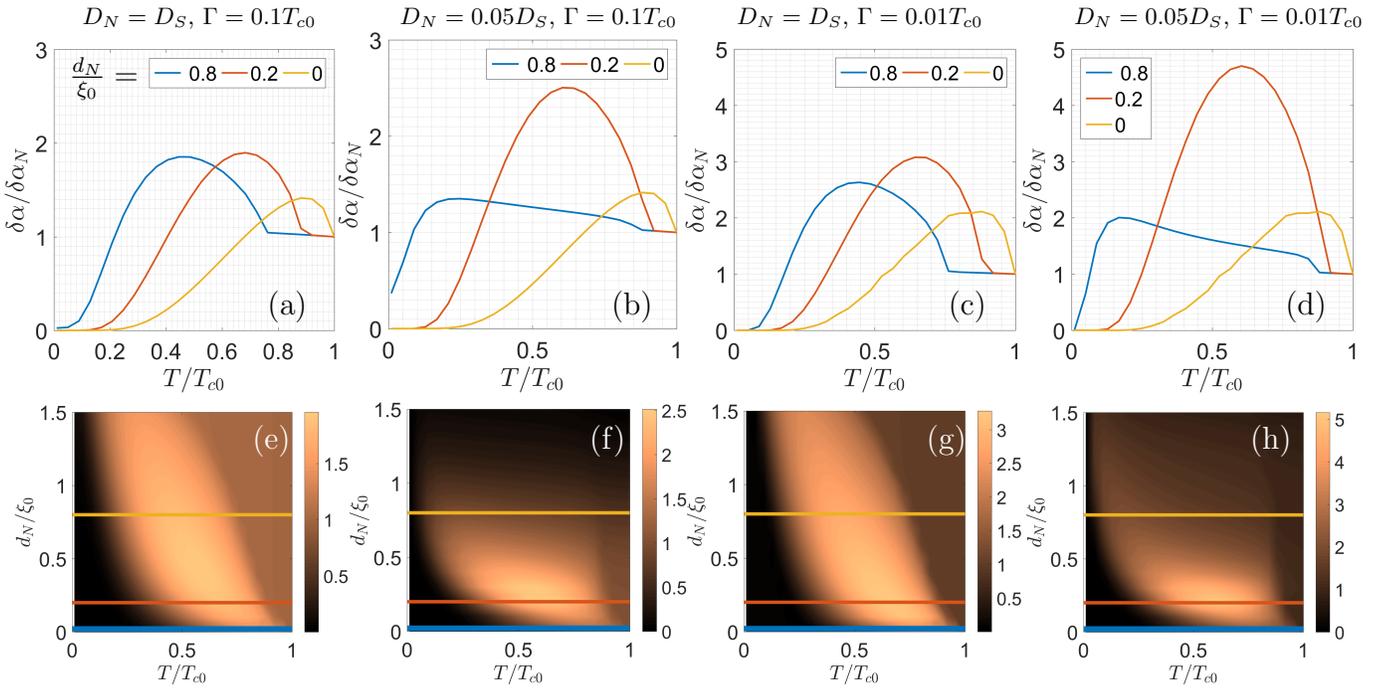
\begin{figure*}[htb!]
 \centerline{
 $  \begin{array}{c}
    \includegraphics[width=0.248\linewidth]
 {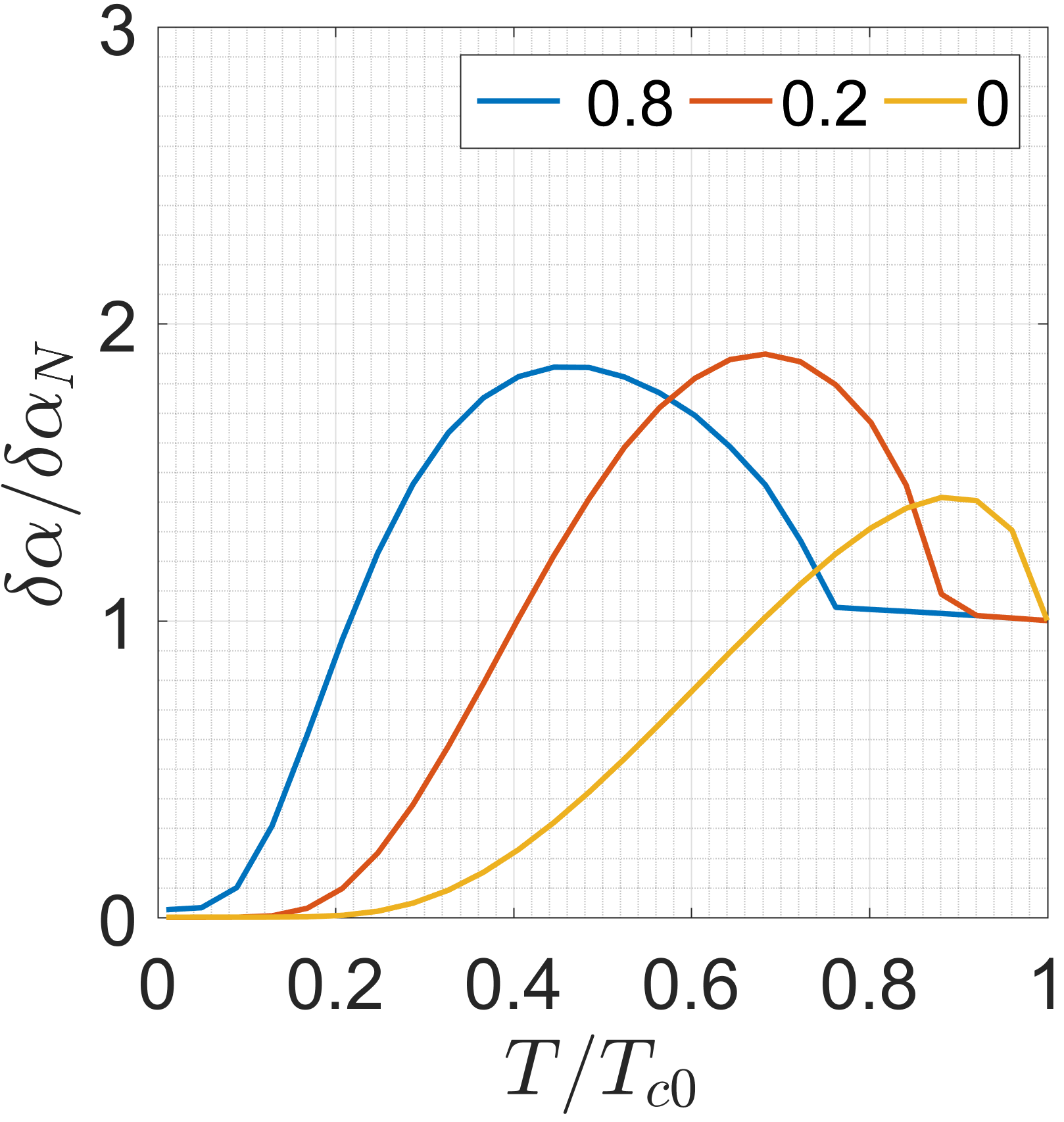} 
 \put (-100,140) { {$D_N=D_S$, $\Gamma =0.1T_{c0}$ } }
   \put (-105,117) { \large {$\frac{d_N}{\xi_0}=$ } }
   \put (-30,30) { \large {(a) } }
   \includegraphics[width=0.255\linewidth]
 {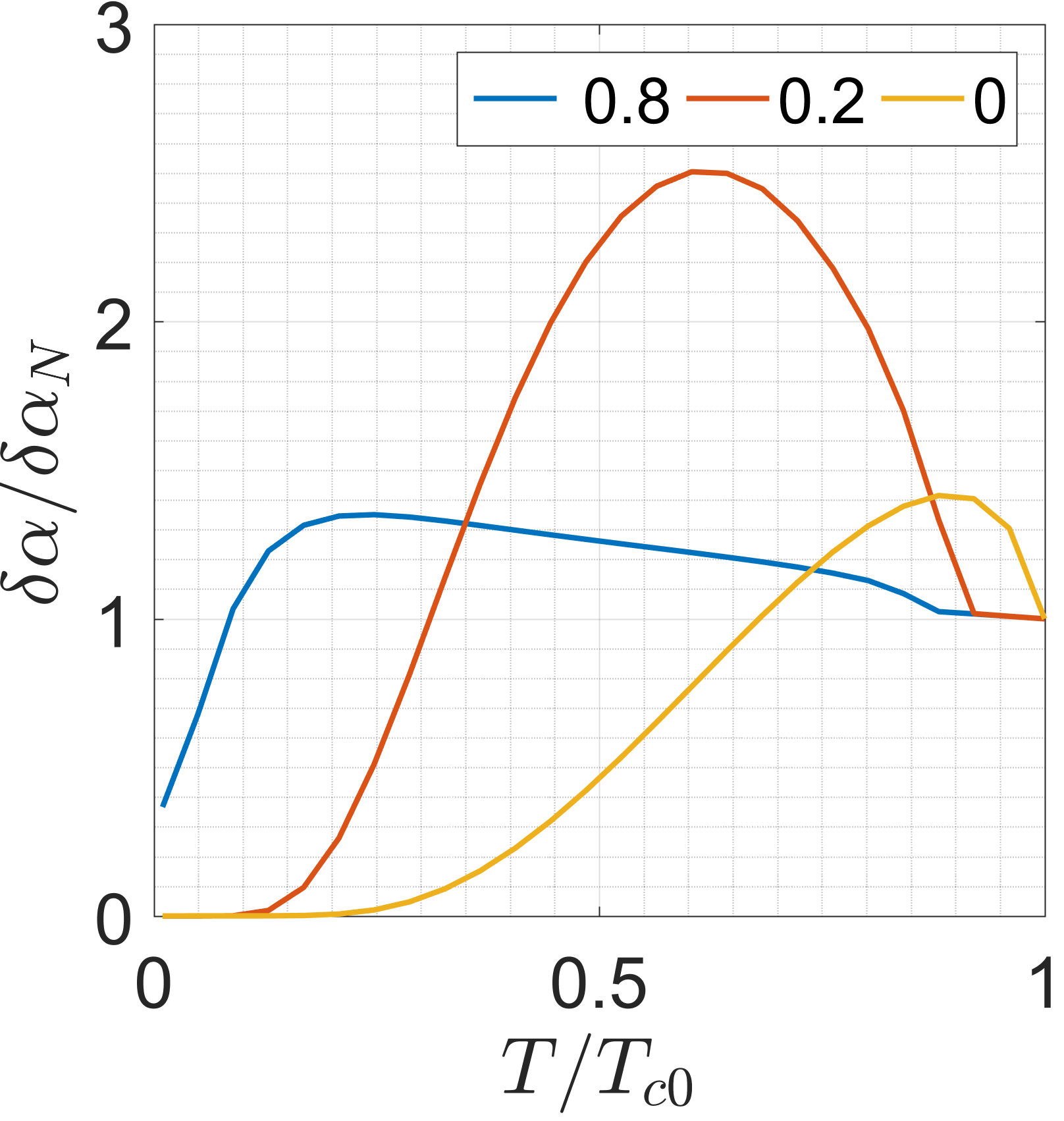} 
  \put (-110,140)  {  {$D_N=0.05D_S$, $\Gamma =0.1T_{c0}$  } }
    \put (-30,30) { \large {(b) } }
     \includegraphics[width=0.250\linewidth]
 {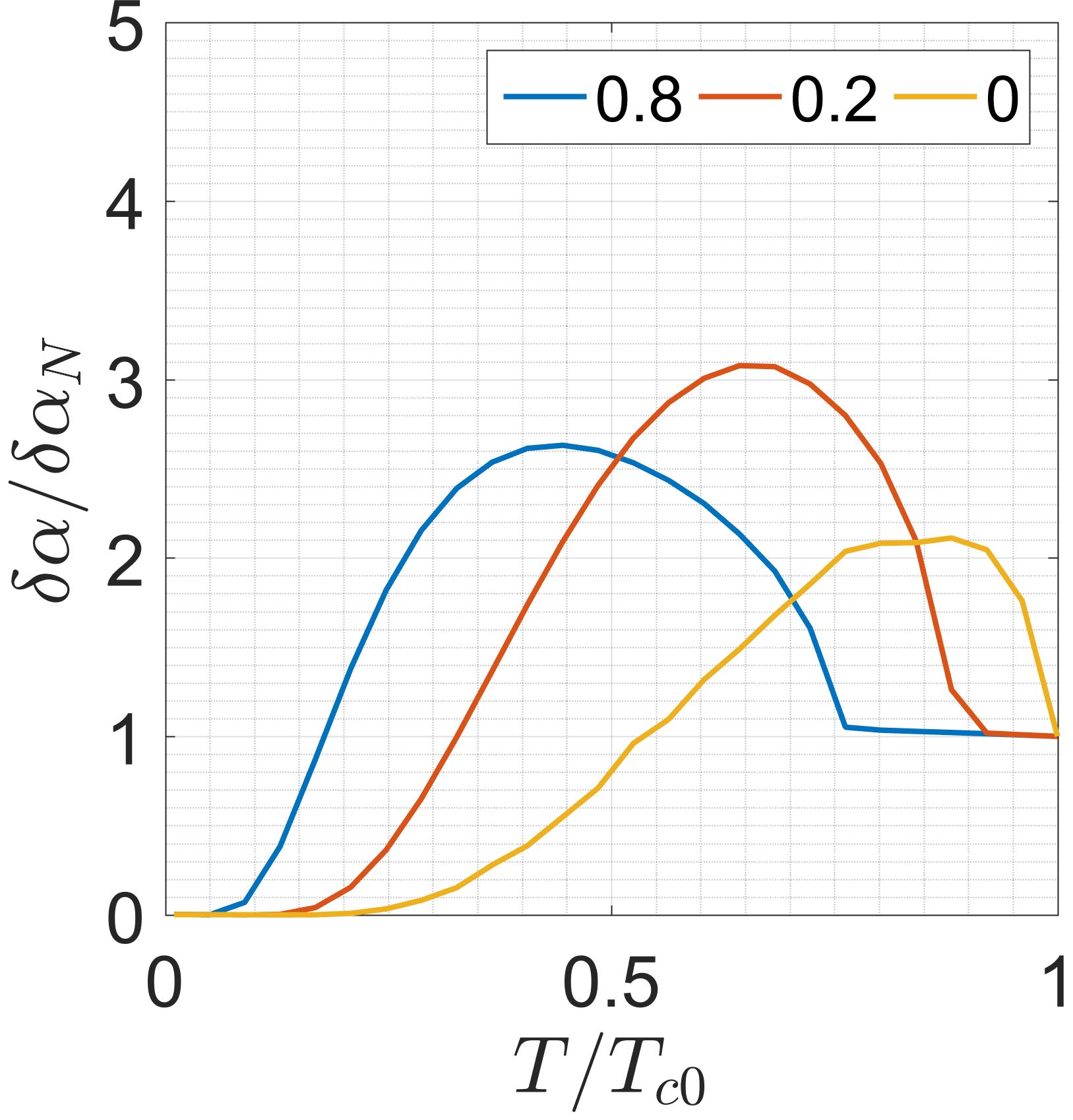} 
 \put (-110,140)  { $D_N=D_S$, $\Gamma =0.01T_{c0}$  }
 \put (-30,30) { \large {(c) } }
   \includegraphics[width=0.250\linewidth]
 {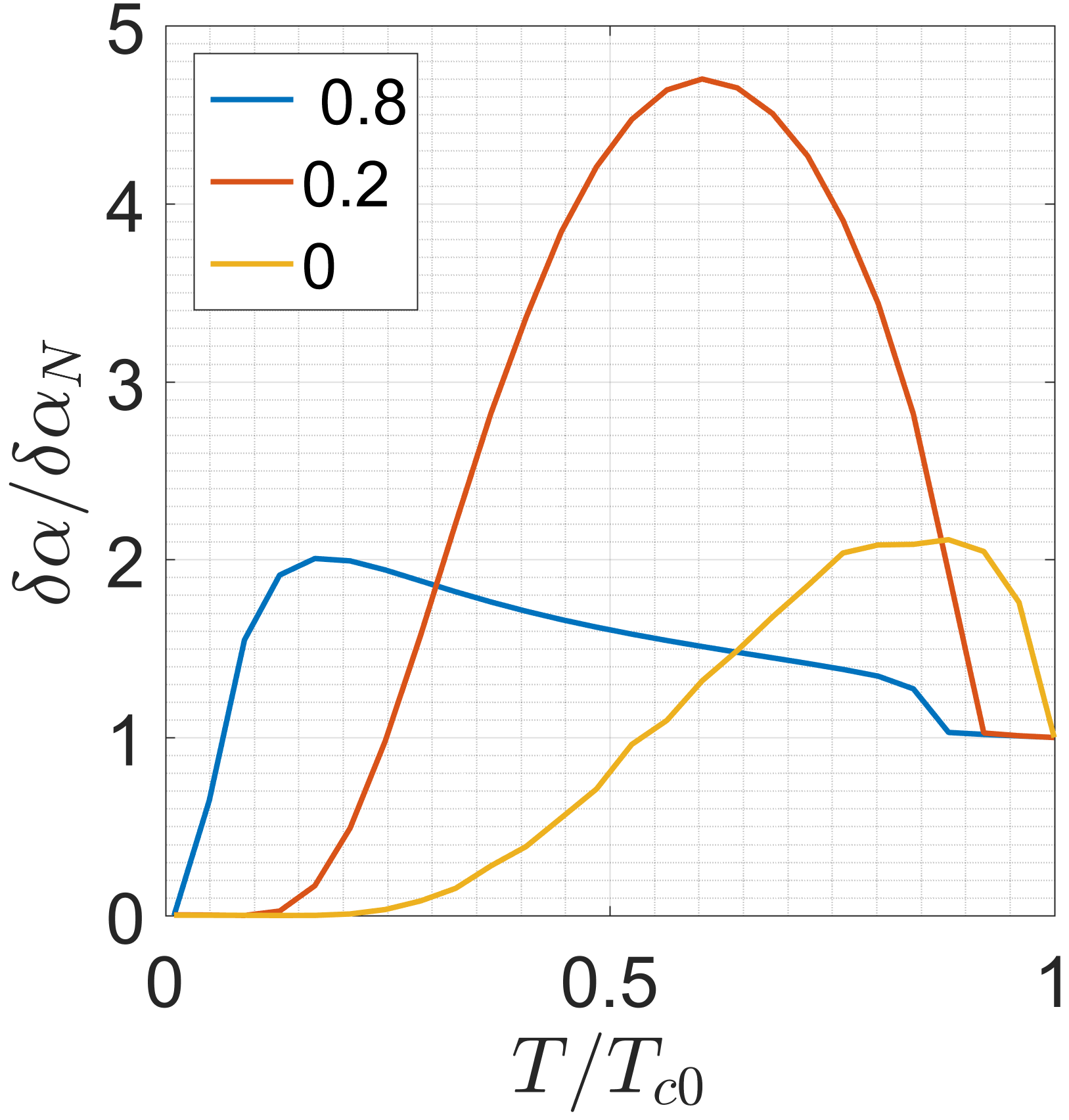} 
   \put (-110,140)  { $D_N=0.05D_S$, $\Gamma =0.01T_{c0}$  }
   \put (-30,30) { \large {(d) } }
  \\
    \includegraphics[width=0.250\linewidth]
 {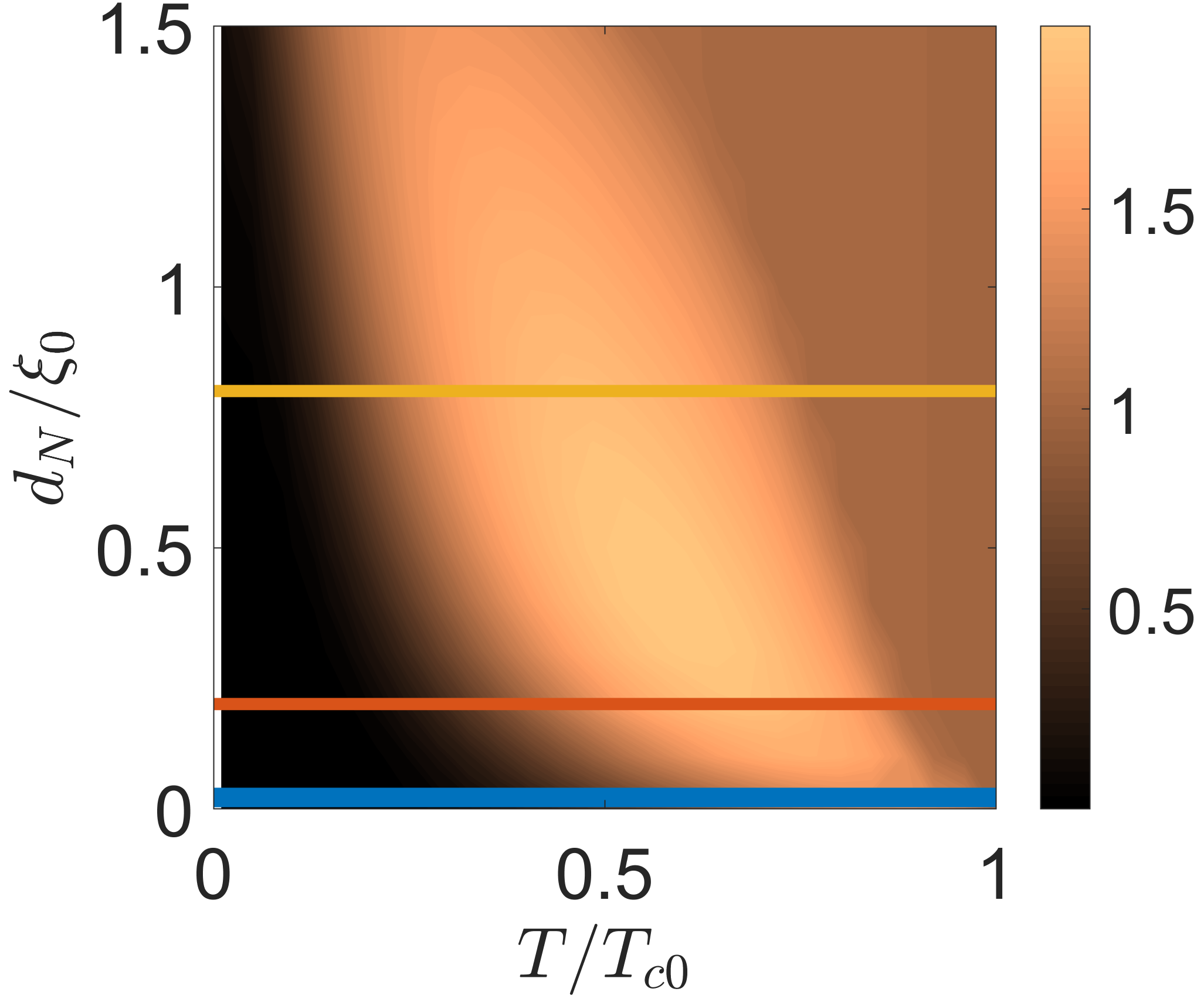} 
 \put (-40,90) { \color{white} \large {(e) } }
 \includegraphics[width=0.250\linewidth]
 {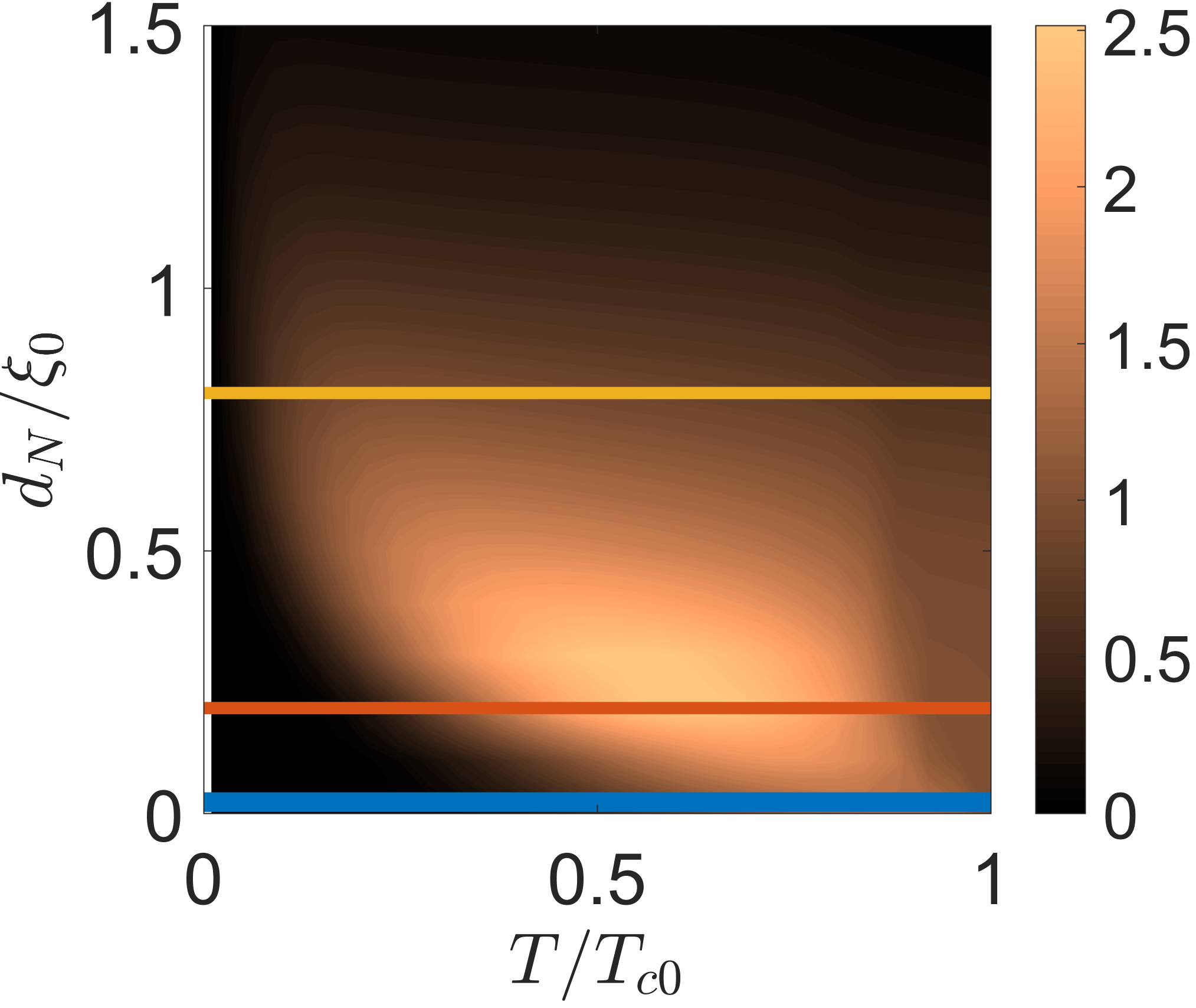} 
 \put (-40,90) { \color{white} \large {(f) } }
 \includegraphics[width=0.250\linewidth]
 {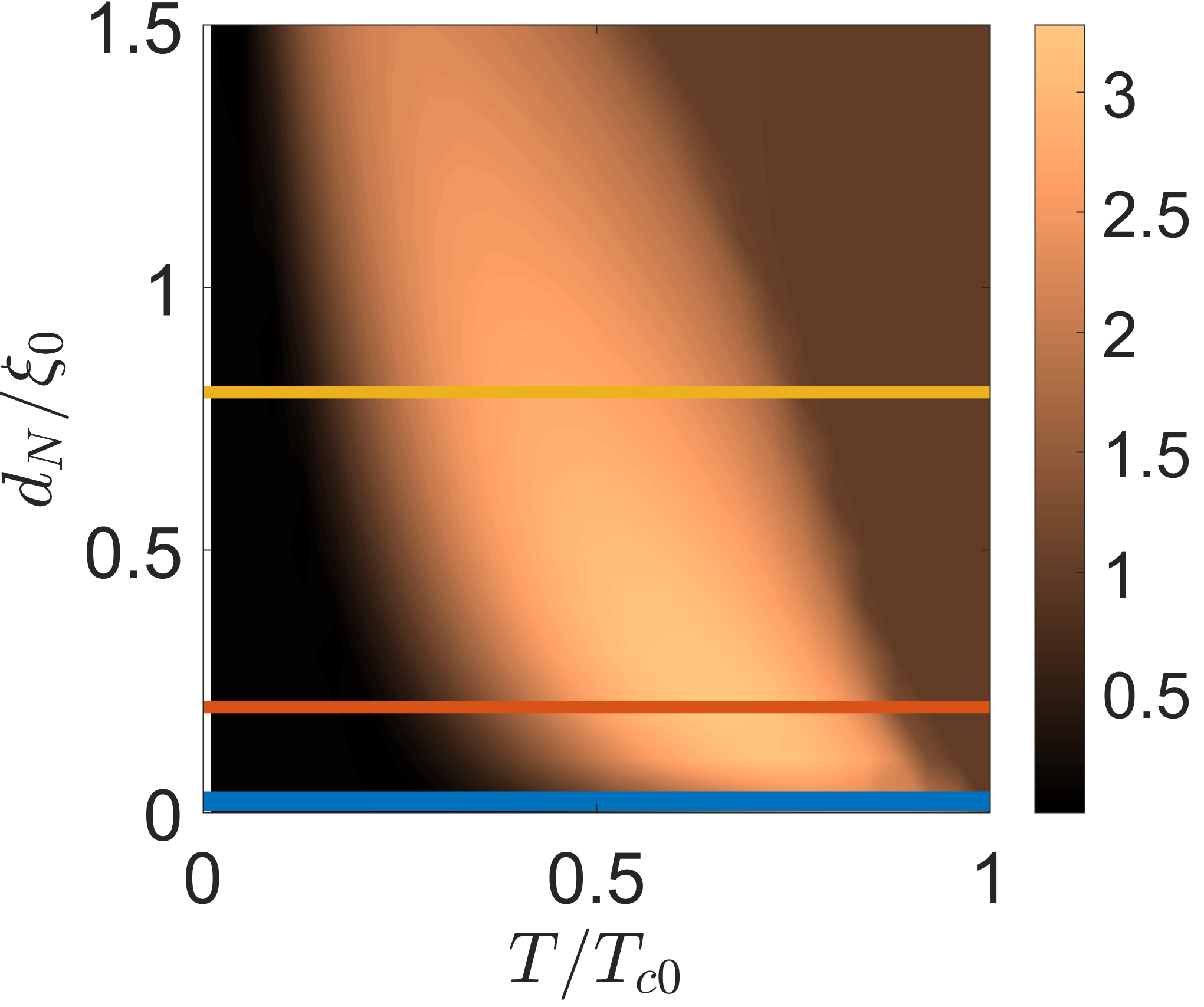} 
 \put (-40,90) { \color{white} \large {(g) } }
 \includegraphics[width=0.240\linewidth]
 {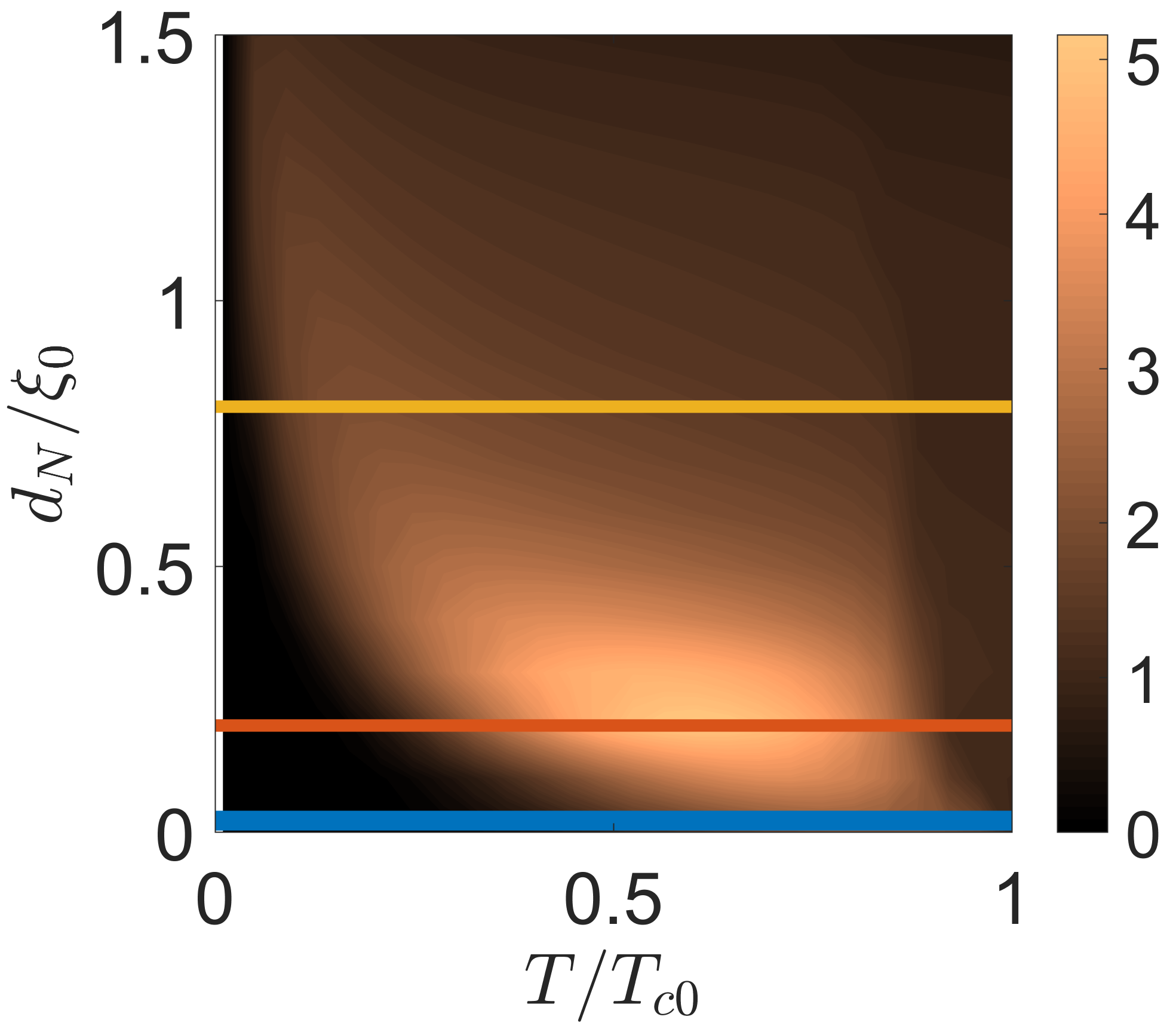} 
 \put (-40,90) { \color{white} \large {(h) } }
  \end{array}$}
  \caption{\label{Fig:3} 
  Upper row:  temperature dependencies of the GD $\delta\alpha (T)$ in FI/N/S systems. The three curves in each plot correspond to  
  $d_N/\xi_0=0.8;\; 0.2; \; 0$. 
  Lower row: color plots of the functions $\delta \alpha (d_N, T) /\delta\alpha_N$. Horizontal lines in each panel are positioned as guide for eyes at
 $d_N/\xi_0=0.8;\; 0.2; \; 0$ corresponding to the curves in the upper plot.  
 The four columns correspond to various Dynes parameters $\Gamma/T_{c0} =0.1;\; 0.01$ and ratios of diffusion coefficients in N and S layers $D_N/D_S = 1;\; 0.05$ specified on top of the panels. 
 Common parameters are 
 $d_S=3\xi_0$, $\tau_{sn}T_{c0} =1$, $\Omega= 0.02 T_{c0}$.    
    }
 \end{figure*}

{\bf Calculation of the time-dependent spin response. }
What is left is to calculate the local spin susceptibility $\chi_m$
 in the Eq.\ref{Eq:GDFiniteThickness} for the FI/N/S structure in Fig.\ref{Fig:1}.  We do so by developing the microscopic kinetic theory of spin pumping  generalizing the quasiclassical approach \cite{millis1988quasiclassical,morten2004spin,Eschrig2015,RevModPhys.90.041001} to the time-dependent situation. 
 
 The magnetization of conduction electrons is determined by  spin accumulation and 
  can be written in terms of the Keldysh quasiclassical Green's function (GF) as 
  \begin{equation} \label{Eq:SpinPolarizationGen}
  \bm S (t) =  - \nu {\rm Tr}\; [\hat\tau_3 {\bm \hat\sigma} g^K (t,t)] /8
 \end{equation}    
    $g^K$ is the (2$\times$2 matrix) Keldysh component of the quasiclassical GF matrix
  $\check{g} = \left(%
 \begin{array}{cc}
  \hat g^R &  \hat g^K \\
  0 &  \hat g^A \\
 \end{array}\label{eq:GF0}
 \right)
$ which depends on two times and a single spatial coordinate variable $\check{g} = \check{g}  (t_1,t_2,{\bm r})$. 
GF $\check g$  obeys the Usadel equation 
 \begin{equation}\label{KeldyshUsadelT}
 \{\hat\tau_3\partial_t, \check g \}_t + \nabla 
 ( D\check g\circ \nabla \check g)  = 
  \Delta [ \hat \tau_1 , \check g ] + [\check \Gamma, \check g] 
 -
 [\check \Sigma_{so}, \check g]_t.
  \end{equation} 
where $\hat\sigma_k,\hat\tau_k$, $k=0,1,2,3$ are Pauli matrices, $D$ is the diffusion coefficient. 
The commutator operator is defined as
   $[X, g]_t= X(t_1) g(t_1,t_2)- g(t_1,t_2) X(t_2)$, similarly for anticommutator 
   $\{,\}_t$. The symbolic product operator is given by
   $ (A\circ B) (t_1,t_2) = \int dt A(t_1,t)B(t,t_2)$. 
   
  Spin relaxation is determined by the spin-orbital  scattering  
  self energy
 \begin{align} \label{Eq:ImpuritySelfEnergy}
  &  \hat\Sigma_{\rm so} ={\bm\sigma}\cdot
  \hat g {\bm\sigma}/(6\tau_{\rm so})
 \end{align}
  The self-consistency equation for the gap function is 
  \begin{align} \label{Eq:SelfConstOP}
  \Delta = \lambda 
  {\rm Tr} [\hat\tau_1 \hat g^K ]/4
  \end{align}
  where $\lambda$ is the pairing coefficient.
  In our model we assume 
  the pairing constant to be suppressed in the N region  
    $\lambda (x<d_N)=0.05 \lambda (x>d_N)$ as compared to its value 
     in S.  
  We  scan over the 
  values of the diffusion coefficient in the N layer $D_N$ while keeping it fixed in S layer $D_S$. 
  The inelastic scattering is described by the Dynes\cite{Dynes1984} parameter which enters to the Eq.\ref{KeldyshUsadelT} as the matrix in Nambu-Keldysh space with $\hat \Gamma^{R,A} = \pm  \Gamma \hat \tau_3 $ 
   which described both the DOS singularity broadening and the relaxation of non-equilibrium distribution functions as described below. Note that this terms conserves the total spin in accordance with the general property of spin-independent electron-phonon scattering \cite{}. 

Eq.\ref{KeldyshUsadelT} is supplemented  by the  {\it dynamical boundary conditions } at $x=0$ describing the 
spin splitting and pumping induced by the electron scattering at the FI interface with time-dependent magnetization.
 These boundary conditions are derived\cite{SupplMat}  from the   spin-dependent  scattering matrix $\hat S$
 connecting the incident $\hat \psi_i$ and reflected $\hat \psi_r$
 electronic waves 
 $ \hat\psi_r = \hat S (t) \hat \psi_i$. 
 For frequencies small compared to the exchange field in FI 
 we use the adiabatic approximation which yields 
 the expression  
  $ \hat S = e^{i (\bm m{\bm \hat\sigma}) \hat\tau_3 \Theta/2 }
 $, where $\Theta$ is the time-independent spin-mixing angle. 
 Then, assuming that $|\Theta|\ll 1$ and 
 \begin{align} \label{Eq:BCFS}
  D \check g \circ \partial_x \check g (x=0) =
  iJ_{sd}[\bm \sigma \bm m \hat\tau_3, \hat g]_t 
 \end{align}
 where $\bm m =  \bm m (t)$ is the time-dependent magnetization.  
 Within the minimal band model of the FI\cite{Tokuyasu1988,millis1988quasiclassical} the interfacial exchange 
 constant is expressed through the spin-mixing angle  as 
   $ J_{sd} = 
  \dfrac{\nu v_F}{4} \int_{-1}^1 d \hat p_{x} |\hat p_{x}| 
  \Theta(\hat  p_{x}) $, where $\hat  p_{x}$ is the electron momentum projection on the interface normal. Eq.\ref{Eq:BCFS} generalizes the 
  static boundary condition at the spin-active interface\cite{Tokuyasu1988,millis1988quasiclassical,Cottet2009,Eschrig2015} to the case of time-pendent magnetization. 
  The induced spin current is obtained using the general expression 
    $
 \bm i (t) = \pi \nu D {\rm Tr} [ \bm{\hat \sigma}  \check g
 \circ \partial_x \check g ](t,t) 
  $. With the help of Eqs.(\ref{Eq:BCFS},\ref{Eq:SpinPolarizationGen}) it yields the phenomenological Eq.(\ref{Eq:SpinCurrentOn}).   

Introducing the usual parametrization of quasiclassical Keldysh function 
   in terms of the distribution function
 $  \hat g^K = \hat g^R\circ \hat f- \hat f\circ \hat g^A
 $ we can identify the terms which are essential to calculate linear response in the low-frequency limit. 
 Expanding the energy representation of $\hat g^K$
to the first order in $\Omega$ we obtain the non-equilibrium correction  
      \begin{align} \label{Eq:LinearResp_gK} 
    & \delta \hat g^K   = 
   (\hat{\bm \sigma} \bm m_\Omega )
   \left[  (\hat g_{0}^R - \hat g_{0}^A ) f_h   + 
   \frac{\Omega \partial_\varepsilon n_0}{2}  
    (  g^R_h +  g^A_h )  \right]
   \end{align} 
   where we parametrise the spin-dependent corrections as follows 
   $\hat f = (\hat{\bm \sigma} \bm m_\Omega )f_h $
 and 
     $\delta g^{R,A}= (\hat{\bm \sigma} \bm m_\Omega ) \delta g^{R,A}_h$. 
     In contrast to stationary non-equilibrium
 situations \cite{morten2004spin} when only the first term in (\ref{Eq:LinearResp_gK}) is important the time-dependent case requires taking into account also the second term with the corrections of spectral functions\cite{2007.04372}. 
In the low-frequency limit the calculation is simplifies by neglecting the frequency dependence of the perturbed spectral GF in (\ref{Eq:LinearResp_gK}). 
 Using (\ref{Eq:LinearResp_gK}) we write the time-dependent spin polarization  in the metallic film as follows  
 \begin{align} \label{Eq:SpinOmega}
  \bm S_\Omega = i \Omega \bm m_\Omega \int_{-\infty}^{\infty} d\varepsilon [ 2 N f_h +  (g_{3h}^R + g_{3h}^A ) \partial_\varepsilon n_0 ]
\end{align}
  where 
 $N = {\rm Tr} (\hat\tau_3\hat g^R)/2$ is the local DOS
  and $g_{3h}^{R,A} = {\rm Tr} (\hat\tau_3\hat g^{R,A}_h)/2 $ .
  Equations for zero-order spectral function $\hat g^{R,A}_0(\varepsilon,x)$, corrections $\hat g^{R,A}_h(\varepsilon,x)$
  and the distribution function $f_h (\varepsilon,\Omega, x)$ 
  are obtained straightforwardly\cite{SupplMat} from Eqs.(\ref{KeldyshUsadelT}, \ref
{Eq:BCFS}). 
 The zero-order GF $\hat g^{R,A}_0(\varepsilon,x)$ are calculated in the N/S structure self-consistently together with the order parameter 
 \ref{Eq:SelfConstOP}. This gives in particular the $\Delta (x)$ and $N(\varepsilon,x)$ profiles shown in Fig.\ref{Fig:1},\ref{Fig:2}. 
 The corrections $f_h$ and $\hat g_{h}^{R,A}$ are determined by the linear equation  \cite{SupplMat}.

  {\bf Results and discussion} 
  Using the described formalism we calculate 
  the non-equilibrium spin polarization (\ref{Eq:SpinOmega}) 
  in the N/S structure shown in Fig.\ref{Fig:1}.   
  This gives us the local susceptibility (\ref{Eq:SOmegaChiM})
  and  the excess GD (\ref{Eq:GDFiniteThickness} ). 
  The resulting temperature dependencies of 
  $\delta \alpha (T)$ are shown in Fig. \ref{Fig:3}
 for various parameters. 
 The first column in Fig.\ref{Fig:3} corresponds to  
$\Gamma=0.1T_{c0}$ and identical diffusion coefficients in 
N ans S layers. In the absence of N layer 
$d_N=0$ there is a usual coherence peak\cite{ } at $T\approx 0.9 T_c$ with the 
small amplitude 
$\delta\alpha/\delta \alpha_N \approx 1.4$.     
Adding the thin N layer with $d_N >0.1 \xi_0$
 leads to the increase of the peak amplitude to $\delta\alpha/\delta \alpha_N \approx 1.9$  and shifting to lower temperatures. 
 
  The peak is enhanced by decreasing the diffusion coefficient $D_N$
  in the normal layer. Qualitatively, this leads to better localization of surface bound states and hence to the increase of surface DOS.  As shown in the second column of Fig.\ref{Fig:3}
  for $D_N=0.05 D_S$ and $\Gamma=0.1T_{c0}$ the peak 
  is enhanced to $\delta\alpha/\delta \alpha_N \approx 2.5$ 
 reached at $T\approx  0.7 T_c$ with $d_N =0.2\xi_0$. 
 This behaviour is quite similar to the experimental observation \cite{yao2018probe}. 
 For larger $d_N>0.5\xi_0 $ the temperature dependence is qualitatively
 changed to the monotonic increase down to the low temperatures.
 As shown by the yellow curve with $d_N =0.8\xi_0$ the increase continues to $T\approx 0.1 T_c$. 
 
 Even larger increase is obtained for smaller Dynes parameters $\Gamma =0.01 T_{c0}$ as shown in the third and fourth columns of the Fig. \ref{Fig:3}. For $D_N=D_S$ we obtain the maximal value $\delta\alpha /\delta\alpha_N = 3$. For $D_N=0.05D_S$ we obtain the maximal value $\delta\alpha /\delta\alpha_N = 4.8$. For all values of $\Gamma$
 we note that for $D_N \ll D_S$ the monotonically increasing $\delta\alpha (T) $ is obtained down to the threshold temperature of the order of Thouless ennergy 
 $T_{th} \approx  D_N/d_N^2 $. 
  As one can see  in the color plots Fig.\ref{Fig:3}f,h for increasing $d_N$ it can be rather small   $T_{th}\ll T_c$.

  The introduced model can explain the observed spin-pumping enhancement  in GdN/NbN system \cite{yao2018probe} assuming that there is a naturally formed thin  normal layer at the FI/S interface. 
 The pairing suppression at the interface can result from various reasons, including 
 magnetic disorder \cite{Abrikosov1961paramagnetic,amato1976measurement}, strong usual disorder\cite{haviland1989onset} or
 the band structure modification \cite{khestanova2018unusual}.
It is straightforward to check our prediction of the enhanced GD by 
fabricating artificial FI/N/S structures with various parameters. 
 
The behaviour of $\delta\alpha (T)$ obtained in Figs.\ref{Fig:3}b,d
 with $d_N=0.8 \xi_0$ is qualitatively similar to the  one observed experimentally  in Py/Nb/Pt
heterostructures  \cite{Jeon2018, jeon2020tunable}. 
In the equilibrium state of our model the spin-triplet superconductivity is absent. Therefore the monotonic increase 
of GD due to the supercondducting correlations 
 is not in principle an  exclusive feature of the system with  spin super-currents. However, the spin-triplet correlations are generated in the non-equilibrium case (\ref{Eq:LinearResp_gK}) providing \cite{2007.04372} significant contribution to the spin response (\ref{Eq:SpinOmega}).  
 
 The developed quasiclassical theory of spin pumping can be generalized to the case of metallic ferromagnets by introducing the 
 finite spin-dependent tunnelling probability through the F/S interface \cite{Bergeret2012a,BergeretVerso2012,Eschrig2015} to the boundary condition (\ref{Eq:BCFS}).
 This provides the way to study  charge and heat transport induced by the magnetic precession as well as  spin torques induced by voltage and temperature biases\cite{zhu2004novel,houzet2008ferromagnetic,holmqvist2012spin, trif2013dynamic, ojajarvi2020nonlinear}.

 {\bf Conclusions}
 We have developed the general formalism to calculate spin-pumping in spatially-inhomogeneous metallic films with spin-active interfaces. 
 As an example we have considered the FI/N/S  structure and found that the the presence of quasiparticle bound states localized near the spin-active interface provides strong enhancement of spin pumping which shows up in the strong increase of the GD coefficient with decreasing temperature below $T_c$. The  model explains large peak of GD in Gd/NbN structures and shows the way to controlling  spin pumping properties in superconducting systems. 
 
 {\bf Acknowledgements}
 This work  was supported by the Academy of Finland (Project No. 297439) and  Russian Science Foundation, Grant No. 19-19-00594. I thank Yakov Fominov for comments. 

\appendix
 

 
 \section{ Stationary spin-mixing scattering matrix}
 
 Near the flat FI/M surface we write wave functions in the form  $\psi_{\bm k_\parallel} e^{i \bm k_\parallel \bm r}$ where $\bm k_\parallel = k_z\bm z + k_y \bm y$  is the conserved momentum parallel to the interface.
  Along $z$ coordinate we have 1D Shrodinger equations
   \begin{align} \label{SMEq:1DShrodingerNonStat}
 &  i\partial_t \psi = (\hat H - \varepsilon_{F\perp})\psi   \\
 & \hat H = -\partial_x^2/2m   +  [\varepsilon_F + V + (\bm m
 { \bm{\hat \sigma}}) V_s] \theta(-x) 
  \end{align}
  where $\bm m = \bm m(t)$.   
   
    Let us first find the frozen scattering matrix which depends adiabatically on time. 
    In this case the energy of incoming and scattered electrons coincide so that writing $\psi \propto e^{i\varepsilon t}$ we get stationary 1D Shrodinger equation
 \begin{align} \label{SMEq:1DShrodinger}
 & \hat H \psi = (\varepsilon + \varepsilon_{F\perp})\psi
 \\
 & \hat H = -\partial_x^2/2m + [\varepsilon_{F} + V_0 + 
 (\bm m { \bm{\hat \sigma}}) V_s] \theta(-x) 
 \end{align}
   where $\varepsilon_{F\perp}= \varepsilon_F - \bm k^2_\parallel/2m$. 
   For the energy we have $\varepsilon = 
   k^2 /2m -\varepsilon_F$ where $k^2 = k_x^2 + k_\parallel^2$. 
   First, we find the scattering matrix writing solutions
 \begin{align}
  & \psi_{k_\parallel} = A_+ e^{i k_x x} + A_- e^{ - i k_x x}
   \\
  &  \psi_{k_\parallel} = B  e^{x/\lambda_\sigma } 
 \end{align}
 where $\lambda_\sigma^{-2} = 
 2m V_\sigma - k_x^2 $
 and $V_{\uparrow (\downarrow)} = V_0 +(-) V_s$ are the spin-up 
 (down) band energies in FI.
 The reflection coefficient $S_\sigma= A_+/A_-$ is then 
 \begin{align}
 S_\sigma = e^{i\varphi} e^{i \sigma \Theta/2} = 
 \frac{1 + i k_x \lambda_\sigma}
 {1 - i k_x \lambda_\sigma  }
 \end{align}
 Since we are interested in spin-dependent reflection phase we get 
 the spin-mixing angle 
 \begin{align} \label{SMEq:SpinMixingAngle}
 e^{i \Theta} = 
 \frac{1 + k_x^2 \lambda_+\lambda_- + i k_x (\lambda_+- \lambda_-)}
 {{1 + k_x^2 \lambda_+\lambda_- - i k_x (\lambda_+- \lambda_-)}  }
 \end{align}
 which yields 
 \begin{align} \label{SMEq:SMangle}
 \Theta/2 = \arcsin\left( 
 \frac{k_x(\lambda_+ - \lambda_-)}{\sqrt{(1+k_x^2\lambda_+\lambda_-)^2 
 + k_x^2(\lambda_+-\lambda_-)^2} }    
 \right)
 \end{align}
 Finally, the spin-dependent part of the scattering matrix 
 connecting the incident $\hat \psi_i$ and reflected $\hat \psi_r$
 electronic waves written in the basis-independent form 
 \begin{align} 
 & \hat\psi_r = \hat S \hat \psi_i 
  \\
 \label{SMEq:Smatrix}
 & \hat S = e^{i (\bm m{\bm \hat\sigma}) \hat\tau_3 \Theta/2 }
 \end{align}

 \section{Time-dependent boundary conditions   at the FI/metal interface}
 
 Here we derive boundary conditions  () starting from the scattering theory of the  interface between FI and metal, either normal or superconducting one. The main difference from the previous works deriving boundary conditions at FI/M interface is that the
 magnetization of FI depends on time $\bm m=\bm m(t)$. 
 
 We consider matrix GF defined in a Keldysh-Nambu-spin space
  \begin{align} \label{smEq:MatsGF}
 \check G (\bm r_1, \bm r_2, t_1,t_2)
 =  
 \begin{pmatrix} \hat G^R & \hat G^K 
 \\ 0 & \hat G^A 
 \end{pmatrix}
  \end{align}
  where retarded, advanced and Keldysh parts are defined in a standard way as follows
  \begin{align}
  & \hat G^R(\bm r_1, \bm r_2, t_1,t_2) = 
   \theta (t_1-t_2) \times 
   \\ \nonumber
  & \left[ \langle \hat\Psi (\bm r_1, t_1) 
  \hat\Psi^+ (\bm r_2, t_2)  \rangle 
  + 
  \langle \hat\Psi (\bm r_1, t_1) 
  \hat\Psi^+ (\bm r_2, t_2)  \rangle
   \right] 
  \\ 
 & \hat G^A (\bm r_1, \bm r_2, t_1,t_2) = 
  \theta (t_2-t_1) \times
  \\ \nonumber
  & \left[ \langle \hat\Psi (\bm r_1, t_1) 
  \hat\Psi^+ (\bm r_2, t_2)  \rangle 
  + 
  \langle \hat\Psi (\bm r_1, t_1) 
  \hat\Psi^+ (\bm r_2, t_2)  \rangle
   \right]   
   \\
  & \hat G^K (\bm r_1, \bm r_2, t_1,t_2) = 
  \\ \nonumber
  & \langle \hat\Psi (\bm r_1, t_1) 
  \hat\Psi^+ (\bm r_2, t_2)  \rangle 
  + 
  \langle \hat\Psi (\bm r_1, t_1) 
  \hat\Psi^+ (\bm r_2, t_2)  \rangle
    \end{align}
 where the field operators 
 $\hat \Psi = 
 ( \hat \psi_\uparrow, \hat \psi_\downarrow, 
 - \hat \psi_\downarrow^+, 
 \hat \psi_\uparrow^+ )$ satisfy the equations of motion  
 \begin{align}
 & i\partial_t \hat\Psi = \hat H (t) \hat \Psi
  \end{align}
 and the Hamiltonian has time-dependent 
 order parameter $\Delta = \Delta (t)$, boundary potential $V=V(t)$ 
 \begin{align}
 \hat H (t) = \hat\tau_3( \bm k^2/2m - \varepsilon_F) + 
 \hat\tau_2 \Delta (t) 
  + \hat V(t)
 \end{align}
 The GF satisfies Gor'kov equations  
 \begin{align}
 &[i\partial_{t_1} - 
 \hat H (t_1, \bm r_1)]\hat G = \delta (t_{12})\delta (\bm r_{12})
 \\
 &\hat G [- i\partial_{t_2} - 
 \hat H (t_2, \bm r_2)] = \delta (t_{12})\delta (\bm r_{12})
 \end{align}
  where $\bm r_{12} = \bm r_1 - \bm r_2$
 and $t_{12}=t_1-t_2$.  
  Assuming the flat FI/M interface we consider 
  transverse momentum components $k_{z,y}$ as conserved quantities.  The perpendicular component $k_x$ changes to the opposite one upon electron reflection. 
  We are interested in the components of GF which are slowly varying as function of the center of mass coordinate $\bm r = (\bm r_1+\bm r_2)/2$ and thus can be written as follows 
 \begin{align} \nonumber
 & \check G_{\bm k_\parallel} ( x_1, x_2, t_1,t_2)=
 \int d\bm r_{12} 
 e^{ - i\bm k_\parallel \bm r_{12} }
 \check G (\bm r_1,\bm r_2,t_1,t_2)
 \end{align}
 The GF satisfies Gor'kov equations 
 \begin{align}
 & [i\partial_{t_1} - \hat H (t_1, z_1) ]
 \check G_{\bm k_\parallel} = \delta (t_{12})\delta (x_{12})
 \\
 & \check G_{\bm k_\parallel}
 [-i\partial_{t_2} - \hat H (t_2, x_2) ]
 = \delta (t_{12})\delta (x_{12})
  \\ \label{SMEq:HamiltonianZ}
 & \hat H (t,x) = 
 - ( \partial_x^2/2m + \varepsilon_{\perp}) 
 \hat\tau_3 + 
 \hat\tau_2 \Delta (t) + \hat V(t)
 \end{align}   
  where 
  $\varepsilon_{\perp} = \varepsilon_{F} - \bm k_\parallel^2/2m $.
  
 Let's consider the Fourier expansion 
 \begin{align} \label{SMEq:FourierExpansion}
 & \check G_{\bm k_\parallel} (x_1,x_2) = \sum_{ k_{1,2}} 
  e^{i(k_1 x_1 - k_2 x_2)} \check G_{\bm k_\parallel}(k_1, k_2) 
 \end{align}
 Near the M/FI interface $z=0$  we can establish the connection between amplitudes 
 \begin{align} \label{SMEq:Gscattering1}
 & \check G_{\bm k_\parallel}(- k_1, k_2) = \hat S (t_1) \check G_{\bm k_\parallel}( k_1<0, k_2)
  \\ \label{SMEq:Gscattering2}
  & \check G_{\bm k_\parallel}( k_1, - k_2) = 
  \check G_{\bm k_\parallel}( k_1, k_2<0) \hat S^+ (t_2)
  \end{align}
 
 From these two relations we get 
 \begin{align}\label{SMEq:Gscattering12}
 \check G_{\bm k_\parallel}( -k_1, - k_2) = 
 \hat S (t_1) \check G_{\bm k_\parallel}( k_1<0, k_2<0)\hat S^+ (t_2)
 \end{align}
 
 Relations (\ref{SMEq:Gscattering1},\ref{SMEq:Gscattering2}) can be obtained as follows. 
 First, consider the vicinity of interface 
 $|x_{1,2}|\ll \xi$ where $\xi= v_x/\Delta$. In this case we can use the simplified equation for GF neglecting the time derivative and order parameter 
 \begin{align}
 & [ ( \partial_{x1}^2/2m + \varepsilon_{\perp}) 
 \hat\tau_3 - \hat V(t_1,x_1)]
  \check G (x_1<x_2)= 0
   \\
&   ( \partial_{x2}^2/2m + \varepsilon_{\perp}) 
 \check G (x_2<x_1) \hat\tau_3 - 
 \check G (x_2<x_1) \hat V(t_2,x_2)]
  = 0
  \end{align}   
 These are two independent equations identical to the Shrodinger equation (\ref{SMEq:1DShrodinger}) at $\varepsilon =0$. Thus 
 we can write the solution 
 \begin{align}
 & \check G_{\bm k_\parallel} (x_1<x_2) = 
 \sum_{k_1>0} [e^{-ik_1x_1}  + e^{ik_1x_1} \hat S (t_1) ] 
 \check F_2(x_2) 
 \\
 & \check G_{\bm k_\parallel} (x_2<x_1) = \sum_{k_2>0} \check F_1(x_1) [e^{ik_2x_2}  + e^{-ik_2x_2} 
 \hat S^+ (t_2) ] 
 \end{align}
 where $\hat F_{1,2}(x) $ in principle can be
 arbitrary functions. 
 Comparing these relations with the general Fourier expansion (\ref{SMEq:FourierExpansion})
 we get Eqs. (\ref{SMEq:Gscattering1},\ref{SMEq:Gscattering2}). 
  
 The quasiclassical GF in general is introduced according to the following general procedure 
 \begin{align} \nonumber
 & \check g_{\bm p} (\bm r) = 
 \frac{1}{\pi}
 \int_{-\infty}^{\infty}
 d \xi_p \int d \bm r_{12}
 e^{-i \bm p \bm r_{12}}
 \hat\tau_3 \check G (\bm r_1,\bm r_2) 
 \end{align}
 Near the flat surface we have only the $z$-dependence 
 \begin{align} \nonumber
  \hat g_{\bm p} (x) =
  \frac{1}{\pi} \int_{-\infty}^{\infty}
 d q e^{-i qx}
 \int_{-\infty}^{\infty}
 d\xi_p \hat\tau_3 \hat G_{\bm k_\parallel}(k_x+q, k_x-q)
 \end{align}
  where we denote $\bm r_{12} = \bm r_1 - \bm r_2$, $\xi_p = (k_z^2 + \bm k_\parallel^2)/2m - \varepsilon_F$. 
  
 Then at $x=0$ we have 
 \begin{align} \nonumber
 & \check g_{\bm{\underline p}} (x=0) 
 = 
 \frac{1}{\pi} \iint_{-\infty}^{\infty}
 d q d\xi_p \hat\tau_3 \check G_{\bm k_\parallel}(- k_x-q, -k_x+q)
 \\
  & \check g_{\bm p} (x=0) =
   \frac{1}{\pi}  \iint_{-\infty}^{\infty}
 d q d\xi_p \hat\tau_3 \check G_{\bm k_\parallel}(k_x+q, k_x-q)
  \end{align}
  Then using relations \ref{SMEq:Gscattering12}
 \begin{align} \nonumber
 & \check g_{\bm p} (x=0) =
 \frac{1}{\pi} \iint_{-\infty}^{\infty} d q d\xi_p 
 \hat\tau_3 
 \check G_{\bm k_\parallel}(k_x+q, k_x-q) =
  \\ \nonumber
  & \frac{1}{\pi} \iint_{-\infty}^{\infty} d q d\xi_p 
 \hat S(t_1) \hat\tau_3 \check G_{\bm k_\parallel}(-k_x-q, -k_x+q)
 \hat S^+(t_2) 
 \approx
 \\ \nonumber
  & \hat S(t_1) 
  \hat g_{\bm{\underline p}} (z=0)
 \hat S^+(t_2) 
 \end{align}
 where in the last relation we assume that $\hat S$
 does not depend on $q$. 
  Finally we get the time-dependent boundary condition for quasiclassical functions
  \begin{align} \label{SMEq:BCquasiclassic}
  \check g_{\bm{\underline p}} (x=0) =
  \hat S(t_1) \check g_{\bm p} (x=0) \hat S^+(t_2) 
  \end{align}
 
  Expanding $\hat S(t) \approx 1 + i \Theta 
  \hat{\bm \sigma} \bm m (t) /2$ in Eq.\ref{SMEq:BCquasiclassic} we get the matrix current at the M/FI boundary 
  \begin{align} \label{SMEq:MatrixCurrentGen1}
   &  \hat I (t_1,t_2) =  
   - \int \frac{d\Omega_p}{4\pi}
    (\bm n \cdot \bm v_F) \check g_{\bm{ p}} (t_1,t_2)  = 
   \\ \nonumber
   & - v_F \int_0^1 d\hat p_x \hat p_z 
   [ \check g_{\bm{ p}} (t_1,t_2) 
   - 
   \check g_{\bm{\underline p}} (t_1,t_2) 
   ]
   \approx 
   \\ \nonumber
  & - \frac{iv_F }{2} \int_0^1 d\hat p_x \hat p_x 
   \Theta(\hat p_x )
   [ \hat{\bm \sigma} \bm m
   \hat\tau_3, \check g ]_t
   \end{align}
  where we denote $[\hat X, \check g]_t= \hat X(t_1)\check g(t_1,t_2) 
  -  \check g(t_1,t_2) \hat X(t_2)$, 
  $\bm n = \bm z$ is the normal to FI interface and denote the
    incident $\hat{\bm{ p}}\cdot\bm n< 0$ and reflected 
     $\hat{\bm{\underline p}}\cdot \bm n> 0$ momenta. 
  
  This expression can be simplified even more if we assume that 
  due to the impurity scattering the anisotropic parts of GF are small. 
  Then we can use two lowest order terms in the spherical harmonics 
  expansion 
  \begin{align}
  \check g_{\bm{ p}} = \check g  +  {\bm p}\cdot\check{\bm g}_a /p 
  \end{align}
  Keeping only the s-wave term we get for the matrix current 
  (\ref{SMEq:MatrixCurrentGen1})
   \begin{align} \label{SMEq:MatrixCurrentGen}
   &  \check I (t_1,t_2) =  
   i \nu^{-1} J_{sd}
   [ \hat\tau_3\hat{\bm \sigma} \bm m, \check g]_t 
   \end{align}
   where the conductance is given by
  \begin{align}
  & J_{sd} = 
  \dfrac{\nu v_F}{4} \int_{-1}^1 d \hat p_{x} |\hat p_{x}| 
  \Theta(\hat  p_{x})     
 \end{align}

   We can find the spin current using the general expression 
    \begin{align}\label{SMEq:SpinCurrentOnShell}
 \bm i (t) = \pi \nu{\rm Tr}_4 [ \bm{\hat \sigma}  \hat I^K (t,t)] .
  \end{align}
  Taking into account  the definition of the  spin density 
   \begin{equation} \label{Eq:SpinPolarizationGenSupp}
  \bm S (t) =  - \nu {\rm Tr}\; [\hat\tau_3 {\bm \hat\sigma} g^K (t,t)] /8
 \end{equation}   
 the spin current  \ref{SMEq:SpinCurrentOnShell} flowing from FI to the spin sink  can be written as 
  \begin{align} \label{SMEq:SpinCurrentOnShell1}
  & \bm i (t)= J_{sd}
  \bm S(t)\times \bm m (t) 
  \end{align}
  
  \section{Equation for the spectral and distribution functions}
   
     {\it Kinetic equation }
 From the Keldysh-Usadel equation in the main text
 we obtain the finite-frequency kinetic equation 
 \begin{align}\label{Eq:KinEq}
 &  \nabla ( {\cal D}  \nabla f_h )
  = 
   [ \tau_{so}^{-1}  + 2 (2\Gamma + i\Omega) N ] f_h  
 \\ \label{Eq:KinEqBC1}
   & {\cal D}  \partial_x f_h (x=0)
  = - 2i h_{eff} N \partial_\varepsilon n_0
  \\ \label{Eq:KinEqBC2}
   &  \partial_x f_h (x=d_S)=0
     \end{align}       
   where ${\cal D} = D {\rm Tr} (1 - \hat g^R\hat g^A)/2$ and
   $\tau_{so}^{-1} =4{\cal D}/3D\tau_{sn}$.
   The system (\ref{Eq:KinEq}, \ref{Eq:KinEqBC1}, \ref{Eq:KinEqBC2})
   is linear with the coefficients determined by the zero-order spectral function. Solving it we find the spin-dependent non-equilibrium distribution function generated by the dynamical spin-active interface.
  
 
     {\it Spectral functions}
  
\begin{figure*}
 \centerline{
 $  \begin{array}{c}
    \includegraphics[width=0.4\linewidth]
  {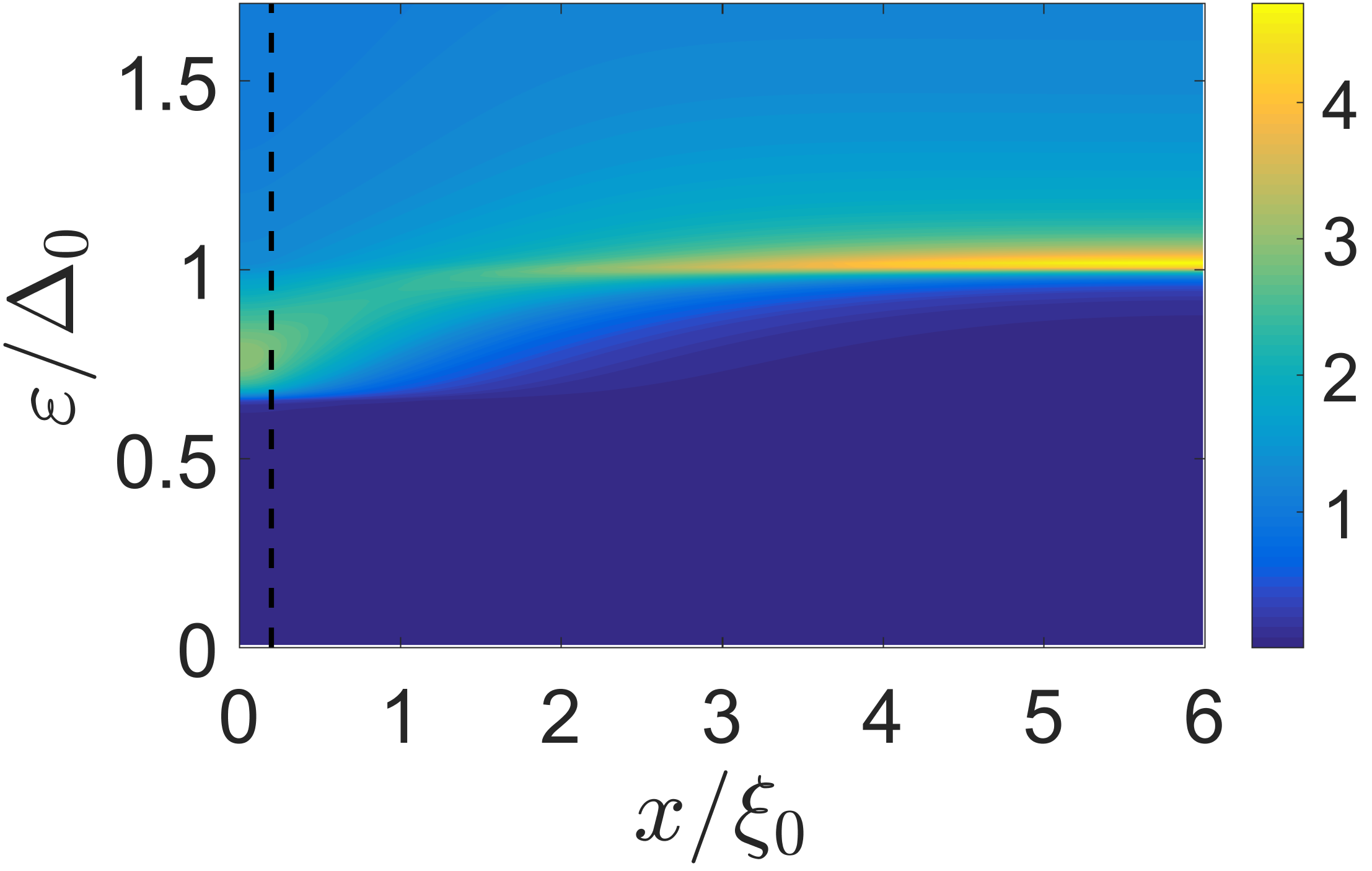}
   \includegraphics[width=0.4\linewidth]
  {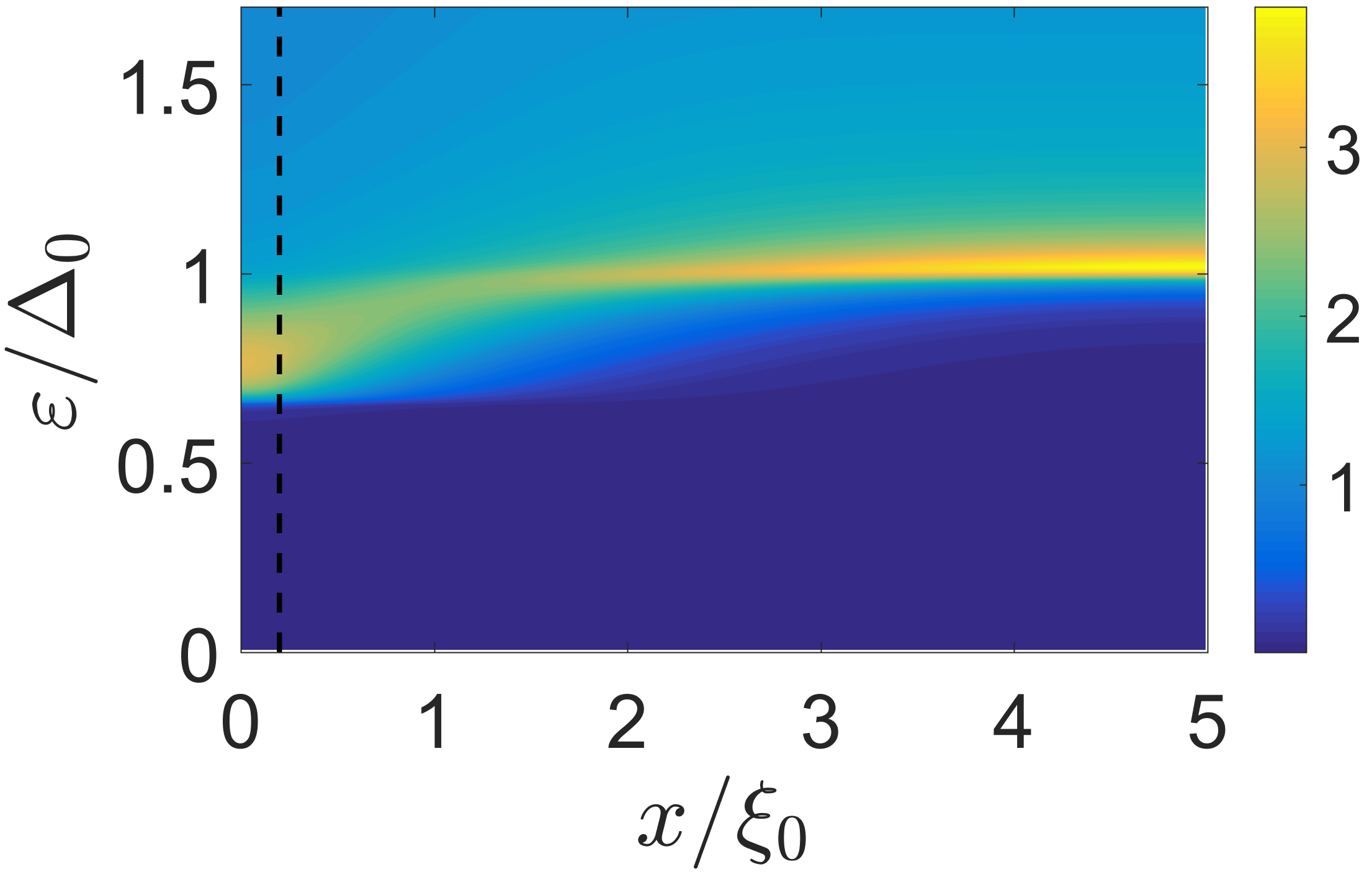}
       \\
      \includegraphics[width=0.4\linewidth]
  {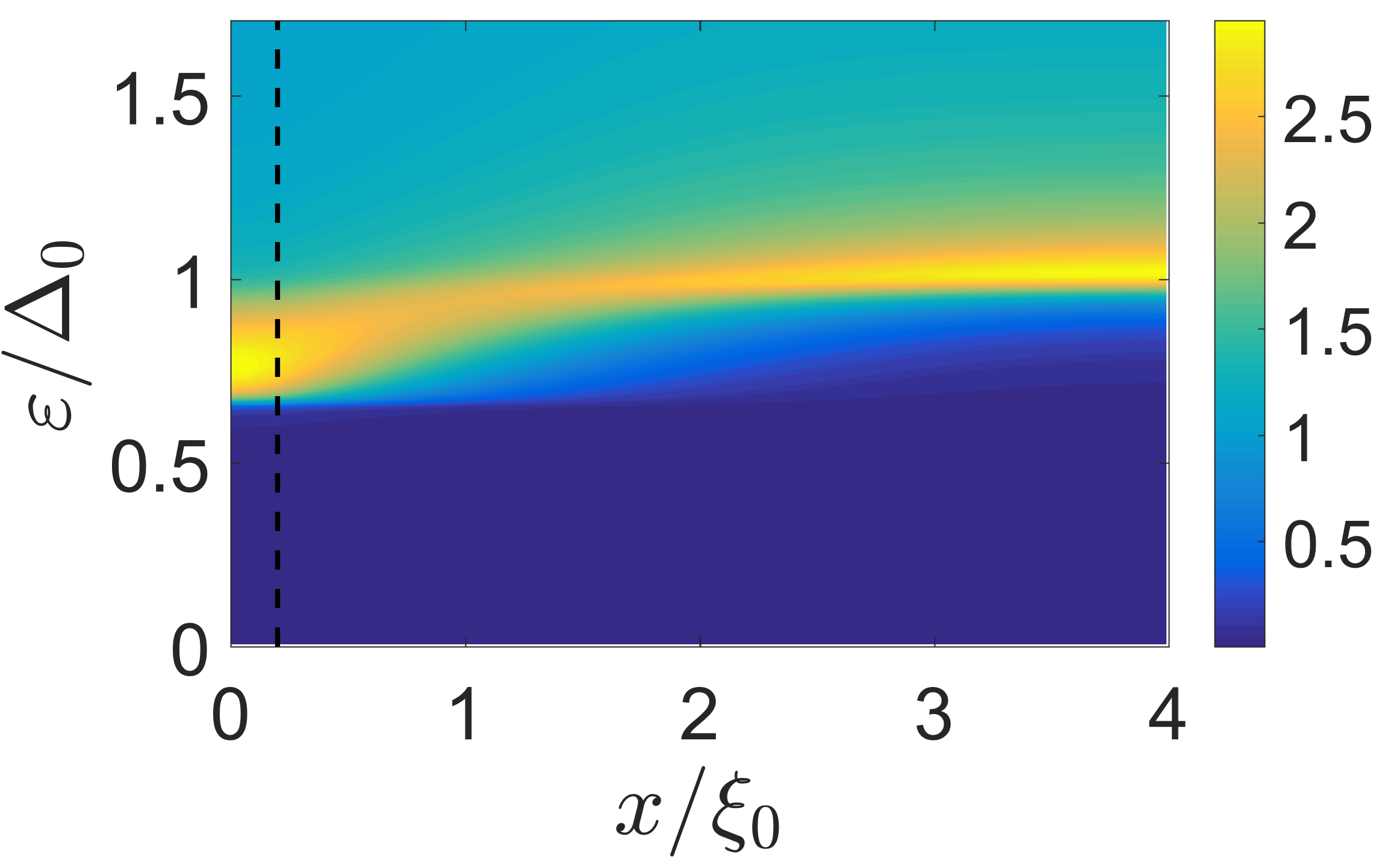}
   \includegraphics[width=0.4\linewidth]
  {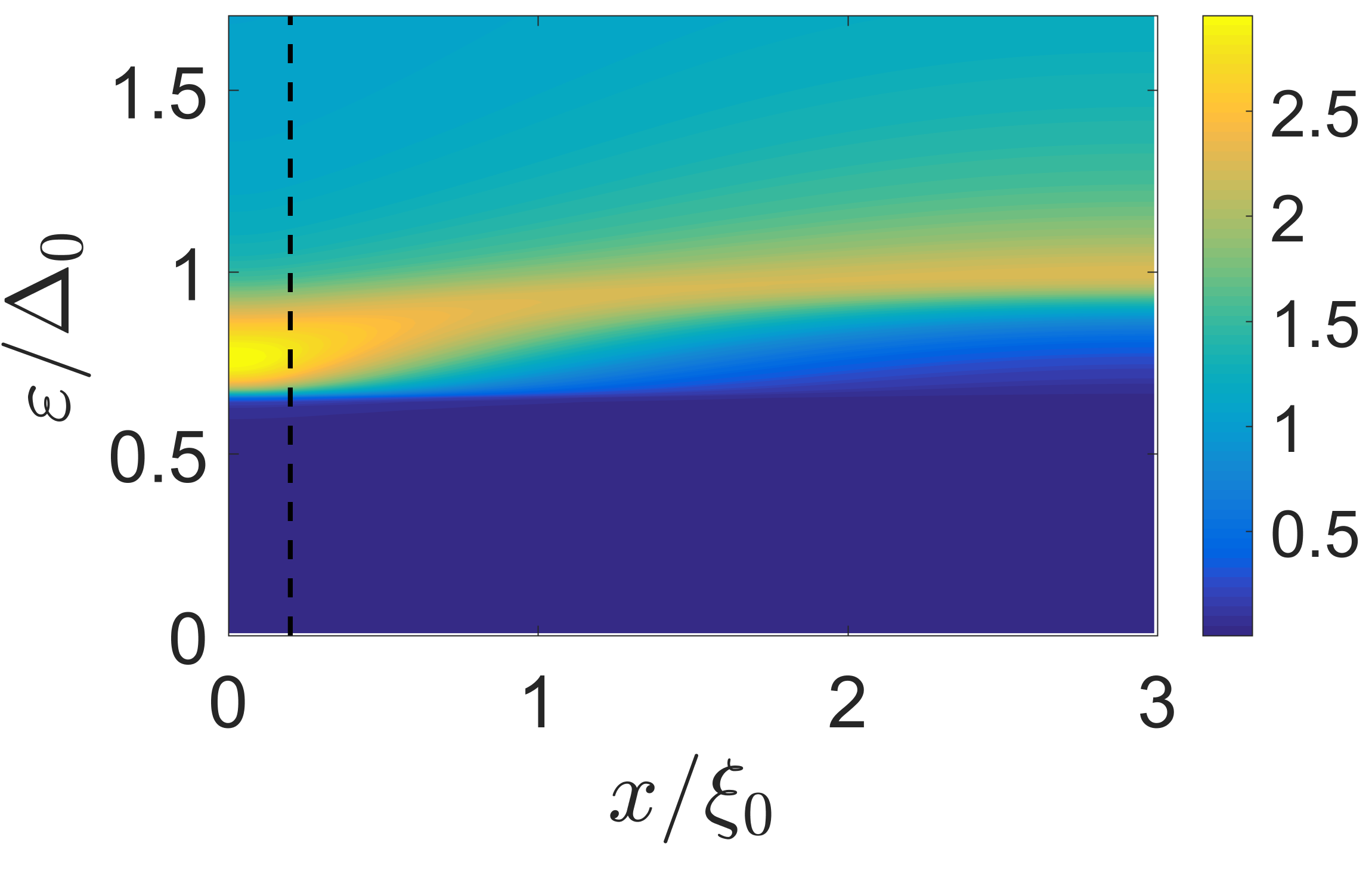}
   \\
   \includegraphics[width=0.4\linewidth]
  {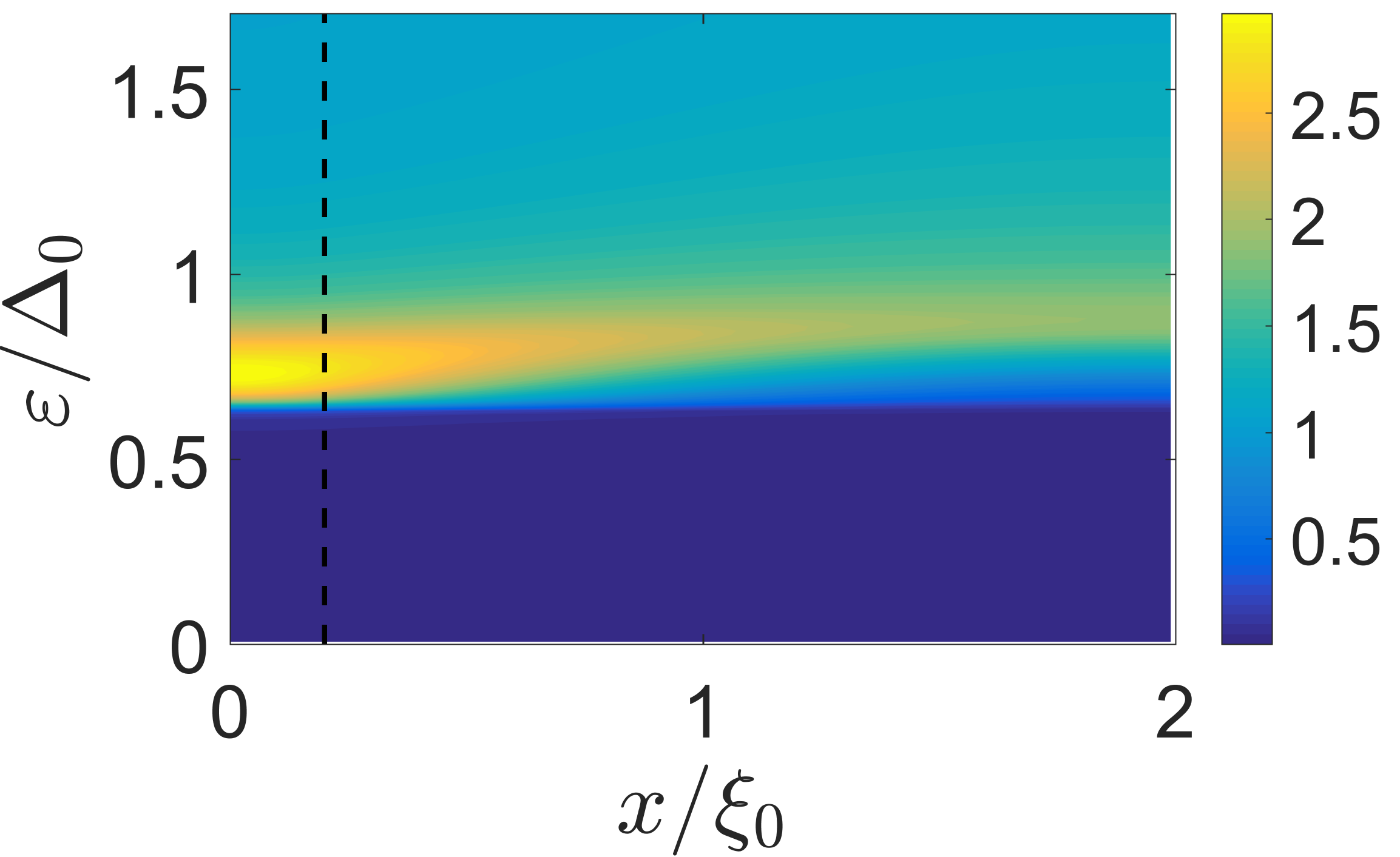}
   \includegraphics[width=0.4\linewidth]
  {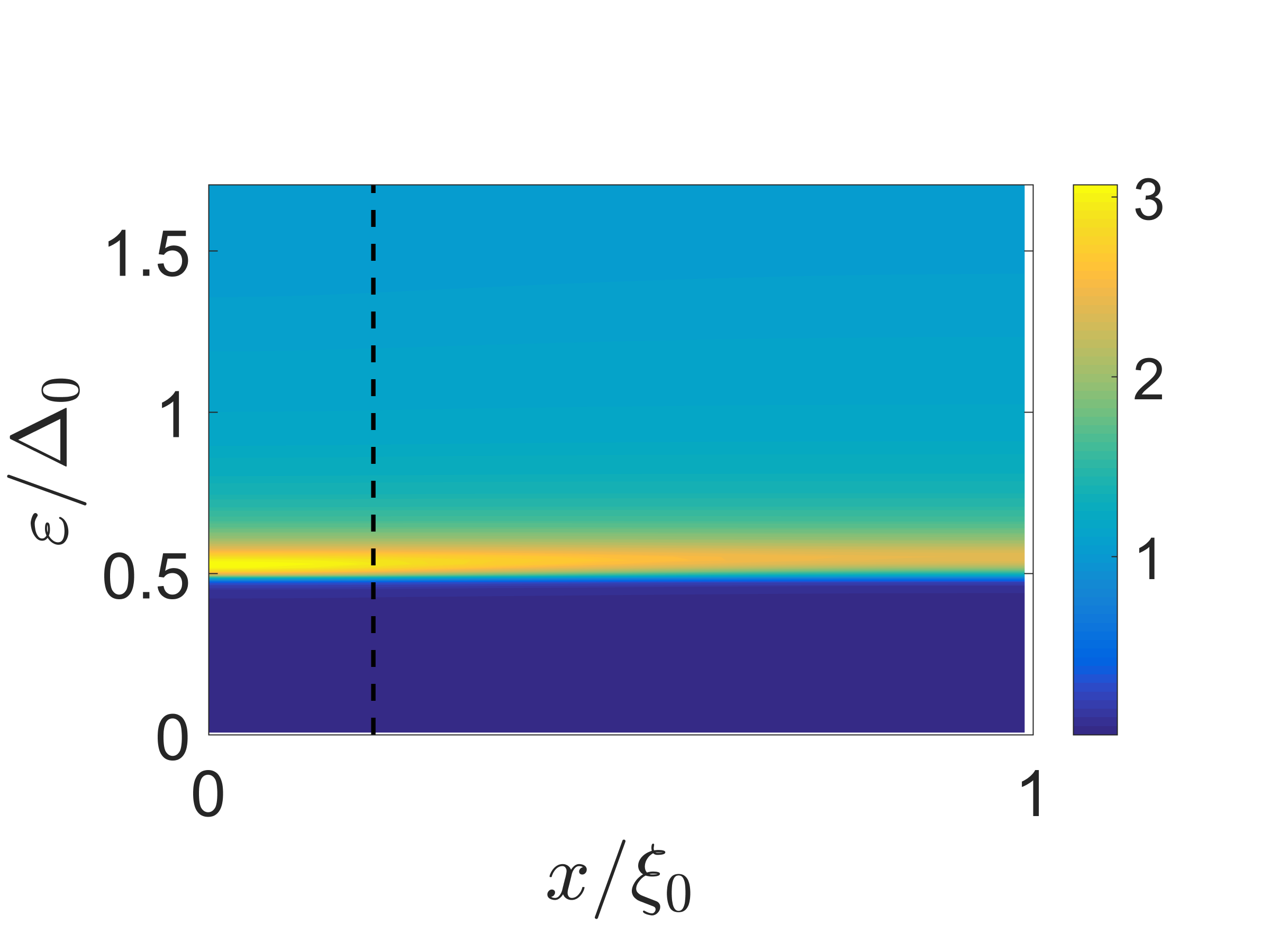}
     \end{array}$}
  \caption{\label{Fig:SM}
  Density of states profile $N(\varepsilon,x)$ in the N/S structures of various lengths. 
The position of N/S boundary shown by the dashed line
 is at $d_N=0.2\xi_0$, $\Gamma=0.01 T_{c0}$, $D_N=D_S$.  }
 \end{figure*}

   In the adiabatic approximation we 
   find the spectral functions from the stationary Usadel equation 
   \begin{equation}\label{Eq:UsadelStat}
    i[(\varepsilon + i \Gamma)\hat\tau_3, \hat g] 
 + 
 \partial_x  
 ( D\hat g \partial_x \hat  g)  = 
  \Delta [ \hat \tau_1 , \hat g ] 
 -
 [\hat \Sigma_{so}, \hat g] 
  \end{equation} 
  with the boundary conditions 
  \begin{align}
 D\hat g\partial_x \hat  g   = iJ_{sd} [ \hat\tau_3\hat{\bm \sigma} \bm m (t), \check g]
\end{align}    
  Using the normalization condition $(\hat g^R)^2=1$ we use the following parametrization for equilibrium GF and
    corrections in the low-frequency adiabatic approximation 
   \begin{align}
 & \hat g^R_{0} = \cos \theta_0 \hat\tau_3 + \sin\theta_0 \hat\tau_1
  \\
 & \hat g^R_h =  ( - \sin  \theta_0 \hat\tau_3 + \cos\theta_0 \hat\tau_1) \theta_h
    \end{align}
   Then we get the following equations for the parameters $\theta_0$, $\theta_h$ 
 \begin{align}\label{Eq:SpectralZeroOrder}
 & i(\varepsilon+i\Gamma) \sin\theta_0
  +\Delta\cos\theta_0 + \partial_x  \left( \frac{D}{2} \partial_x \theta_0 \right) =
    0  
 \\ \label{Eq:SpectralZeroOrderBC}
  & \partial_x \theta_0  (x=0,d_S )=0
  \end{align}
\begin{align} \label{Eq:SpectralCorrection}
 & \theta_h \left[ i (\varepsilon+i\Gamma) \cos\theta_0 -
  \frac{2}{3\tau_{so}}  -
 \Delta  \sin\theta_0 \right] +  \partial_x \left( \frac{D}{2} \partial_x \theta_h  \right) =
    0
 \\ \label{Eq:SpectralCorrectionBC1}
 &  D_N\partial_x\theta_h  (x=0)= 
  2i h_{eff} \sin\theta_0; 
  \\ \label{Eq:SpectralCorrectionBC2}
 & 
  \partial_x\theta_h  (x=d_S)= 0 
 \end{align}
  Solving the nonlinear Eq.(\ref{Eq:SpectralZeroOrder},\ref{Eq:SpectralZeroOrderBC}) together with the self-consistency equation for $\Delta$ we obtain the zero-order spectral functions in the N/S structure. The corresponding DOS profiles are shown in Fig.\ref{Fig:2} and in more detail in Fig.\ref{Fig:SM}.  
  Using them we find the coefficients in the linear Eq.(\ref{Eq:SpectralCorrection},\ref{Eq:SpectralCorrectionBC1},\ref{Eq:SpectralCorrectionBC2}) for the correction   $\theta_h$ which yields the perturbation of spectral functions by the spin-active interface. 
  

 \bibliography{refs2} 

 \end{document}